\documentclass[prb,twocolumn,showpacs,preprintnumbers,superscriptaddress,amsmath,amssymb]{revtex4}

\usepackage{graphicx}
\usepackage{dcolumn}
\usepackage{bm}
\usepackage{epstopdf}
\usepackage{xcolor}
\usepackage[normalem]{ulem}

\begin{document}


\title{Tuning of quantum interference in top-gated graphene on SiC}

\author{Andrea~Iagallo}
\affiliation{NEST, Istituto Nanoscienze--CNR and Scuola Normale Superiore, \\Piazza San Silvestro 12, 56127 Pisa,  Italy}

\author{Shinichi~Tanabe}
\affiliation{NTT Basic Research Laboratories, NTT Corporation, 3-1 Morinosato Wakamiya, Atsugi, Kanagawa, 243-0198, Japan}

\author{Stefano~Roddaro}
\affiliation{NEST, Istituto Nanoscienze--CNR and Scuola Normale Superiore, \\Piazza San Silvestro 12, 56127 Pisa,  Italy}
\affiliation{Istituto Officina dei Materiali CNR, Laboratorio TASC, 34149 Trieste, Italy}

\author{Makoto~Takamura}
\affiliation{NTT Basic Research Laboratories, NTT Corporation, 3-1 Morinosato Wakamiya, Atsugi, Kanagawa, 243-0198, Japan}

\author{Hiroki~Hibino}
\affiliation{NTT Basic Research Laboratories, NTT Corporation, 3-1 Morinosato Wakamiya, Atsugi, Kanagawa, 243-0198, Japan}

\author{Stefan~Heun}
\email{stefan.heun@nano.cnr.it}
\affiliation{NEST, Istituto Nanoscienze--CNR and Scuola Normale Superiore, \\Piazza San Silvestro 12, 56127 Pisa,  Italy}



\date{\today}

\begin{abstract}
We report on quantum-interference measurements in top-gated Hall bars of monolayer graphene epitaxially grown on the Si face of SiC, in which the transition from negative to positive magnetoresistance was achieved varying temperature and charge density. We perform a systematic study of the quantum corrections to the magnetoresistance due to quantum interference of quasiparticles and electron-electron interaction. We analyze the contribution of the different scattering mechanisms affecting the magnetotransport in the $-2.0 \times 10^{10}$~cm$^{-2}$ to $3.75 \times 10^{11}$~cm$^{-2}$ density region and find a significant influence of the charge density on the intravalley scattering time. Furthermore, we observe a modulation of the electron-electron interaction with charge density not accounted for by present theory. Our results clarify the role of quantum transport in SiC-based devices, which will be relevant in the development of a graphene-based technology for coherent electronics.
\end{abstract}

\pacs{72.80.Vp, 73.20.Fz, 73.43.Qt}
\maketitle

\section{\label{secIntro} Introduction}

In the last decade, graphene emerged as a promising material for a variety of technological applications\cite{NovoselovRMP2011} largely thanks to its intrinsic two-dimensionality. In microelectronics, for instance, progress was recently made in high-frequency\cite{LinScience2010,BourzacNat2012} and metrology\cite{TzalenchukNatNano2010} based applications. From a fundamental point of view, graphene is an exciting material for its unique electronic properties, notably its linear energy spectrum and the chiral nature of its charge carriers. In particular, chirality manifests at quantizing fields with the characteristic half-integer quantum Hall effect, a clear signature of monolayer graphene. 

At low temperatures, the low--field magnetoresistance of 2D conductors can be affected by quantum interference.\cite{AltshulerPRB1980} This phenomenon originates from the dynamics of counterpropagating quasiparticles along autointersecting orbits, when phase coherence is retained. In conventional conductors, owing to the time-reversal symmetry of this process, quasiparticles interfere constructively at the origin. This leads to an enhanced backscattering probability, which in turn produces an increased zero-field resistance. This effect is called Weak Localization (WL).\cite{PhysRevLett.42.673,HikamiPTP1980,BergmannPRB1983,BergmannPR1984} Conversely, when destructive interference occurs, backscattering is suppressed, thus zero-field resistance decreases, and Weak Anti Localization (WAL) is observed. In conventional conductors, WAL is due to spin-orbit interaction or scattering at magnetic impurities.\cite{BergmannPR1984,HikamiPTP1980}

The peculiar electronic properties of graphene give rise to unusual quantum-interference effects. Owing to the chirality of graphene carriers, a quasiparticle propagating along an autointersecting path acquires an additional (Berry) phase of $\pi$,\cite{NovoselovNat2005} which leads to destructive interference. Since spin-orbit interaction is weak in graphene,\cite{HuertasPRB2006} WAL provides therefore reliable evidence of charge-carrier chirality.

Quantum-interference phenomenology in graphene is therefore driven by the interplay between chirality and different types of elastic-scattering events: their relative weight determines the regime of localization observed (WL or WAL). Trigonal warping of the energy dispersion is known to suppress chirality, thus strongly suppressing WAL.\cite{McCannPRL2006} The presence of smooth potential variations (as produced, e.g., by ripples or remote impurities in the substrate) is also known to reduce quantum interference.\cite{MorpurgoPRL2006,FalkoSSC2007} The cumulative effect of these $intra$valley chirality-breaking mechanisms on the transport properties is accounted for by a characteristic elastic scattering time $\tau_{*}$. 

On the other hand, interaction of quasiparticles with atomically-sharp defects (such as missing atoms or device edges) was recently linked to $inter$valley scattering, which causes carriers to change abruptly the Dirac valley. While phase-coherence can be preserved in this process, the memory of the chirality is lost, with the final result of restoring WL. The strength of $inter$valley scattering can be quantified in terms of the $inter$valley elastic-scattering time $\tau_{iv}$. Localization effects are observed as long as quasiparticles maintain their phase coherence, i.e., within the timescale of the inelastic dephasing time $\tau_{\varphi}$ and as long as $\tau_{\varphi}>\tau_{iv},\tau_{*}$.

A finite magnetic field breaks time reversal symmetry by adding an extra phase to quasiparticles that propagate along closed paths. When a large number of different paths is present, the effect of interference is averaged to zero, and resistance recovers its classical value $R_{0}$. This confines interference effects to a narrow field range around $B=0$, where both WL and WAL can contribute to the magnetoresistance. The suppression of interference by a magnetic field is rather dramatic, and leads to a sharp peak (dip) in the magnetoresistance that is the signature of WL (WAL). 

The interference phenomena described so far are ultimately single-particle effects and involve non-interacting quasiparticles. Electron-Electron Interaction (EEI) between carriers can lead to an additional quantum correction to magnetoresistance,\cite{AAEEI} which can in principle be observed even in the presence of magnetic-field intensities that would suppress quantum interference. This contribution stems from Coulomb scattering between quasiparticles, which is strongly enhanced in the presence of disorder owing to longer interaction times. The effects of EEI were extensively studied over the last decades in conventional 2D systems,\cite{LiPRL2003,GornyiPRB2004,BockhornPRB2011} and only recently in graphene.\cite{Kozikov2010,JouaultPRB2011,Lara-AvilaPRL2011,JobstPRL2012}

Detailed investigations on quantum interference and EEI were reported on mechanically ex\-foliated\cite{TikhonenkoPRL2009,TikhonenkoPRL2008} and quasi-freestanding\cite{JobstPRL2012} graphene. On the other hand, results on these effects with epitaxial graphene on SiC are limited. In particular, the interplay between localization and chirality is still largely unexplored for this type of graphene and only positive magnetoresistance was observed so far.\cite{Lara-AvilaPRL2011,BakerPRB2012} 
Carrier-density changes are expected to modulate EEI strength, but their effect on quantum interference is not trivial. Experimental investigations on exfoliated graphene showed the carrier-density dependence of the scattering mechanisms affecting localization in graphene, which can also be used to drive the crossover from WL to WAL regime.\cite{TikhonenkoPRL2009,TikhonenkoPRL2008} The difficulty to realize conventional backgating delayed the investigation of quantum interference in epitaxial graphene at different density regimes, and only photochemical-gated devices could be investigated so far.\cite{BakerPRB2012}

The purpose of this paper is a detailed investigation of quantum interference and EEI contributions to the low-field magnetoresistance in top-gated epitaxial graphene. We performed a systematic characterization of the transport properties as a funcion of temperature and carrier density. For the measurements, we made use of a top-gate electrode to change the charge density in our device and were able to tune both the quantum interference and EEI effects in the low-field magnetoresistance. In particular, we report an evolution from negative to positive magnetoresistance as temperature is increased and density is lowered. We shall show that our results are well described by the current theory of localization in graphene\cite{McCannPRL2006,FalkoSSC2007} and EEI in 2D conductors,\cite{AAEEI,GirvinPRB1982} and that the degree of disorder in our epitaxial graphene device is comparable to that in high-quality exfoliated graphene. 

\section{\label{secExpDetails} Experimental Details}

\begin{figure}[tbp]
\includegraphics[width=\columnwidth]{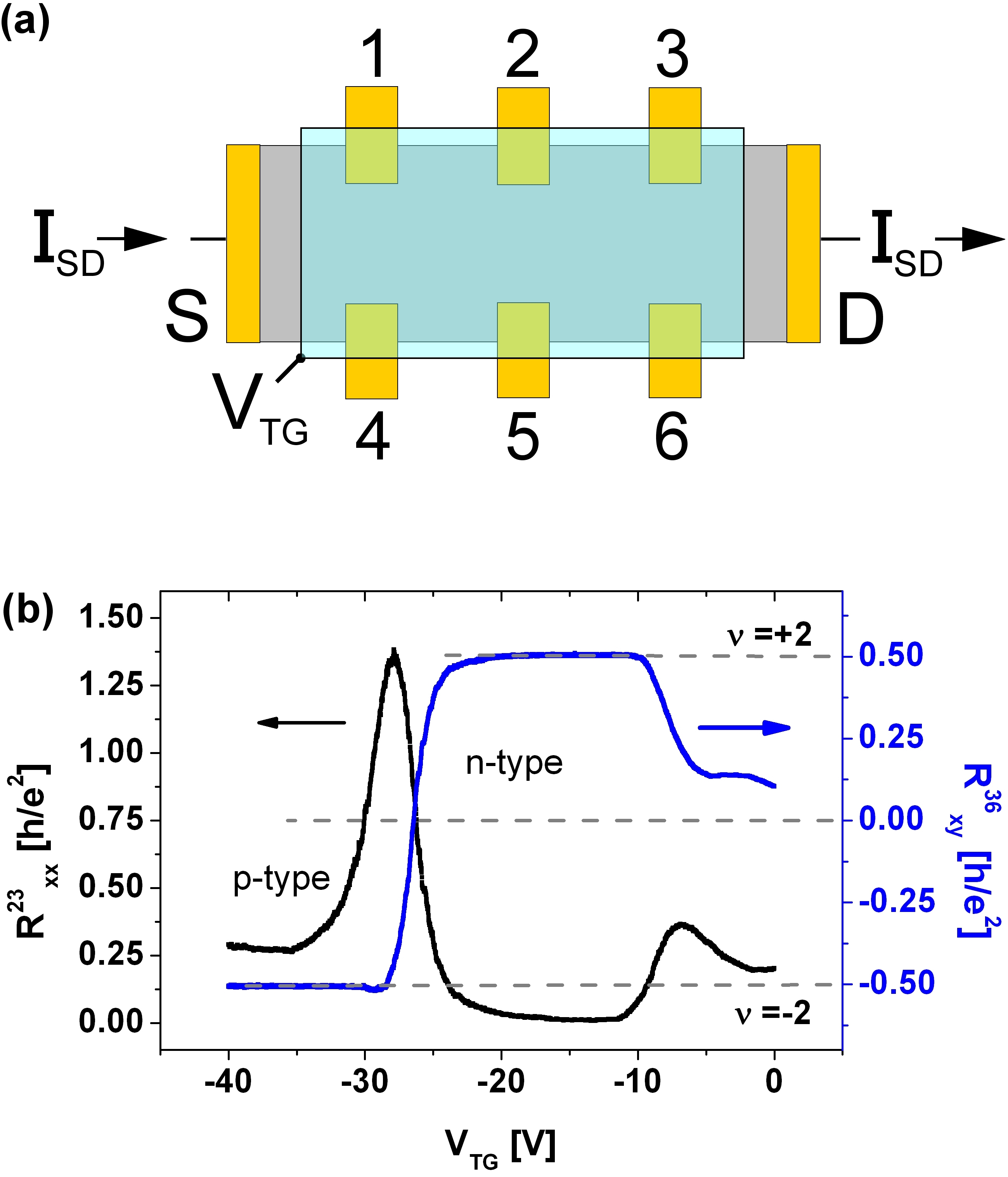}
\caption{(Color online) (a) Schematic of our device and measurement setup. Grey: graphene, blue: top gate, yellow: ohmic contacts. The longitudinal and transversal resistances, $R^{23}_{xx}=V^{23}_{xx}/I_{SD}$ and $R^{36}_{xy}=V^{36}_{xy}/I_{SD}$, respectively, are recorded while the charge density in the device is independently set by the application of a top-gate voltage $V_{TG}$.  (b) Half-integer quantum Hall effect measured at T=250 mK and $B=1.6$~T, clearly showing plateaus in $R^{36}_{xy}$ at filling factor $\nu= \pm 2$ and a change of carrier polarity at $V_{TG}= V_{CNP}\approx -27$ V.}
\label{fig1}
\end{figure}

Devices analyzed in this work are large-area graphene Hall bars (length~$\times$~width~$=$~300~$\mu$m~$\times$~50~$\mu$m) fabricated by standard optical lithography from an epitaxial graphene layer grown on a SiC(0001) wafer. Hall bars were further processed to pattern Cr/Au (5/250~nm) metallic contacts and to deposit a bilayer dielectric.\cite{TanabeAPEX2010} The dielectric consists of 140~nm of Hydrogen Silsequioxane (HSQ) and 40~nm of SiO$_{2}$, spin-coated and sputtered onto the substrate, respectively. Finally, a large area Cr/Au (10/180~nm) top gate was defined by e-beam lithography. Figure~\ref{fig1}(a) shows a sketch of our device. Magnetotransport measurements were performed by standard lock-in technique in a Heliox He$^{3}$ cryostat with a base temperature of 250~mK. The longitudinal and transversal resistances $R^{23}_{xx}=V^{23}_{xx}/I_{SD}$ and $R^{36}_{xy}=V^{36}_{xy}/I_{SD}$, respectively, were measured in a 4-point configuration. A bias current $I_{SD}=10$~~nA was used to avoid overheating of the device, and voltages up to $V_{TG}=-45 $~~V were applied to the top gate to tune the charge density. Thanks to the large dimension of the Hall bar, WL and WAL features are easily recognizable since they are not masked by universal conductance fluctuations that can affect $\mu$m-sized bars.\cite{TikhonenkoPRL2008} In particular, averaging of the measured resistance over large ranges of $V_{TG}$ was not necessary. 

Magnetic fields in the $0-11$ T range were used to characterize the device in the Quantum Hall (QH) regime. The device displayed the Half-integer QH effect as shown in Fig.~\ref{fig1}(b). This is a fingerprint of monolayer graphene, since multilayer graphitic systems are known to produce a QH effect with plateaus in Hall resistance at standard integer positions.\cite{NovoselovNat2005,Novoselov2006} The change of sign of $R^{36}_{xy}$ reflects the change of the carrier polarity occurring at the Charge Neutrality Point (CNP), as determined by applying a top-gate voltage $V_{TG}=V_{CNP}\approx -27$ V. The position of the CNP determined by the maximum in $R^{23}_{xx}$ is shifted towards more negative values of $V_{TG}$ because of a slight anisotropy in the carrier concentration. This feature, discussed in the next paragraph, was already observed in epitaxially grown devices.\cite{TanabePRB2011} The onset of the $2e^{2}/h$ quantized plateau in $R^{36}_{xy}$ was reached at different magnetic field values depending on $V_{TG}$, occurring at $\left|B\right|>5$~T for $V_{TG}=0 $~V. 

In order to compensate for the small charge-density inhomogeneity observed, in the following we shall show ${R_{xx}=\frac{1}{2}(R^{23}_{xx}+R^{56}_{xx})}$. This inhomogeneity stems from the fabrication technology chosen to form the dielectric layer: spin-coating deposition, required for HSQ, results in thickness variations of $1-3\%$ over a $10-100$~$\mu$m length scale, as estimated by AFM measurements. This variation leads to a charge density gradient in the graphene layer, which is known to affect the experimental data curves.\cite{PonomarenkoSSComm2004} For linear density variations in the longitudinal direction, the best resistance estimate is obtained by averaging the values measured at opposite sides of the device.\cite{KarmakarPhysE2004} The residual small tilt in the data of Fig.~\ref{fig3}, observed at all temperatures and for all $V_{TG}$, is due to the nonlinear part of the density gradient, and was corrected by processing our data as described in detail in Appendix~\ref{APP}.

\begin{figure}[tbp]
\includegraphics[width=\columnwidth]{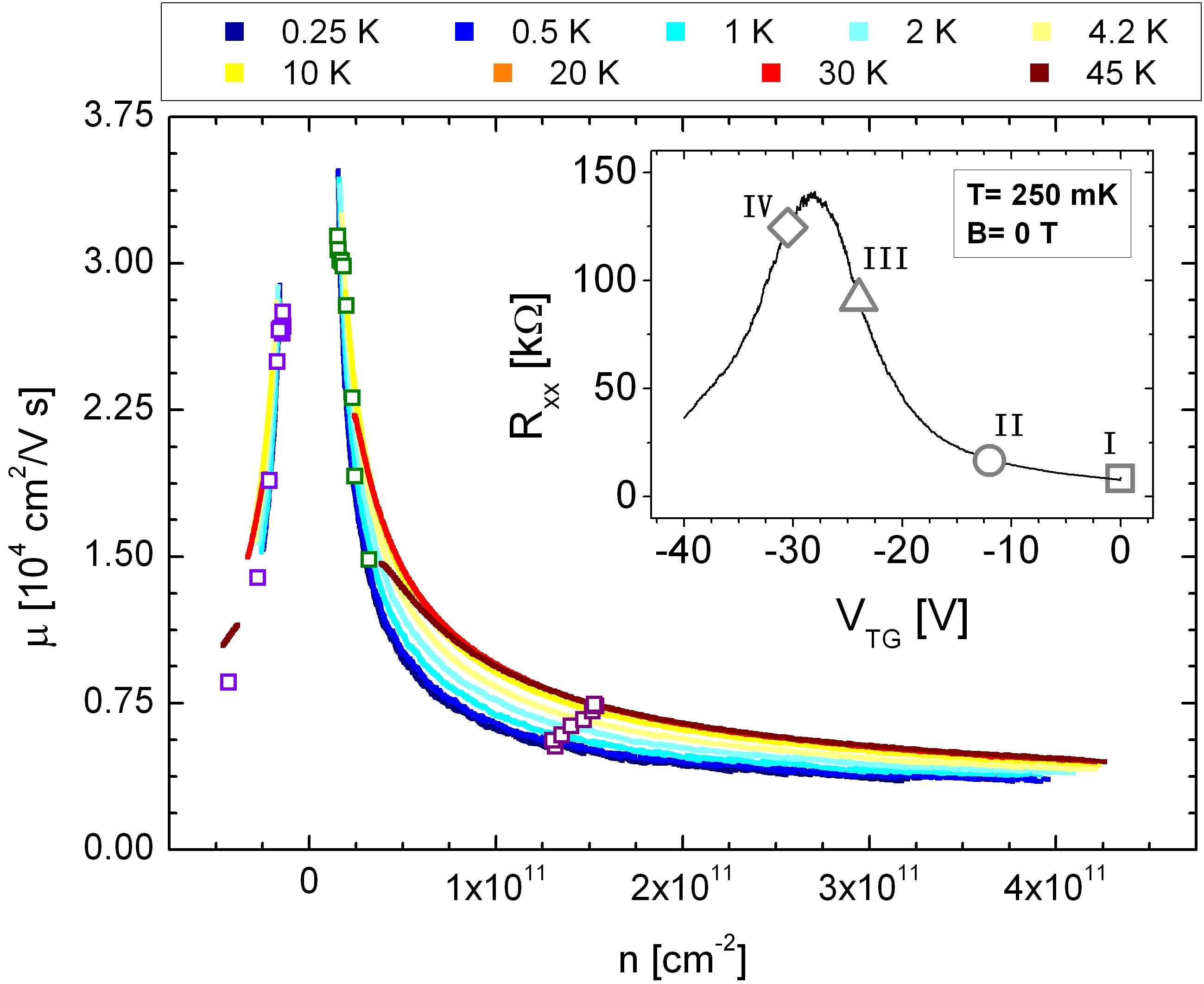}
\caption{(Color online) Dependence of mobility $\mu$ on carrier density $n$ obtained in the temperature range $0.25-45$~K with the procedure described in the text. $\mu-n$ points obtained from $R_{xx}(0)$ and $R_{xy}(B)$ are shown as empty squares. Inset: zero-field longitudinal resistance $R_{xx}$ as a function of $V_{TG}$ measured at $T = 250$~mK. The four values of $V_{TG}$ selected to study the magnetoresistance are indicated by symbols. The corresponding carrier density values are: I~($3.75 \times 10^{11}$~cm$^{-2}$), II~($1.43 \times 10^{11}$~cm$^{-2}$), III~($2.02 \times 10^{10}$~cm$^{-2}$), and IV~($-2.03 \times 10^{10}$~cm$^{-2}$).}
\label{fig2}
\end{figure}

Before extracting the individual contributions of quantum interference and EEI to the magnetoresistance, we estimate the mobility $\mu$ as a function of charge density $n$ following Ref.~\onlinecite{TanabePRB2011}. For this purpose, the charge density was obtained from measurements of $R_{xy}$ as a function of $V_{TG}$ at $B=\pm 0.2$~T, while the mobility was calculated from the resistivity $\rho$ as $\mu=1/n e \rho$. By cross-correlating the two results, we obtain the $\mu-n$ diagrams shown in Fig.~\ref{fig2} for different temperatures in the 0.25-45 K range. As a comparison, the values of $n$ and $\mu$ estimated from fits of $R_{xy}$ and from $R_{xx}(0)$ are also shown in the same figure. The results obtained with the two methods are in good agreement.

The longitudinal resistance $R_{xx}$ is shown in the inset of Fig.~\ref{fig2} for $B=0$~T and $T=250$~mK and displays a peak at the CNP. The resistance maximum occurs at $V_{TG} \approx -27$~V, in agreement with the value obtained from the QH measurement of $R^{36}_{xy}$ in Fig.~\ref{fig1}(b). In the next Sections, we investigate the low-field corrections to magnetoresistance measured at different top-gate voltages $V_{TG}$. The selected values correspond to significant variations of charge density across the CNP, with four values ranging from $3.75 \times 10^{11}$~cm$^{-2}$ to $-2.0 \times 10^{10}$~cm$^{-2}$, where the negative signs stands for the change of the carrier polarity occurring at the CNP. The density values chosen in this work are displayed in the inset of Fig.~\ref{fig2}.

\section{\label{secExpRes} Results and Discussion}

\begin{figure}[tbp]
\includegraphics[width=\columnwidth]{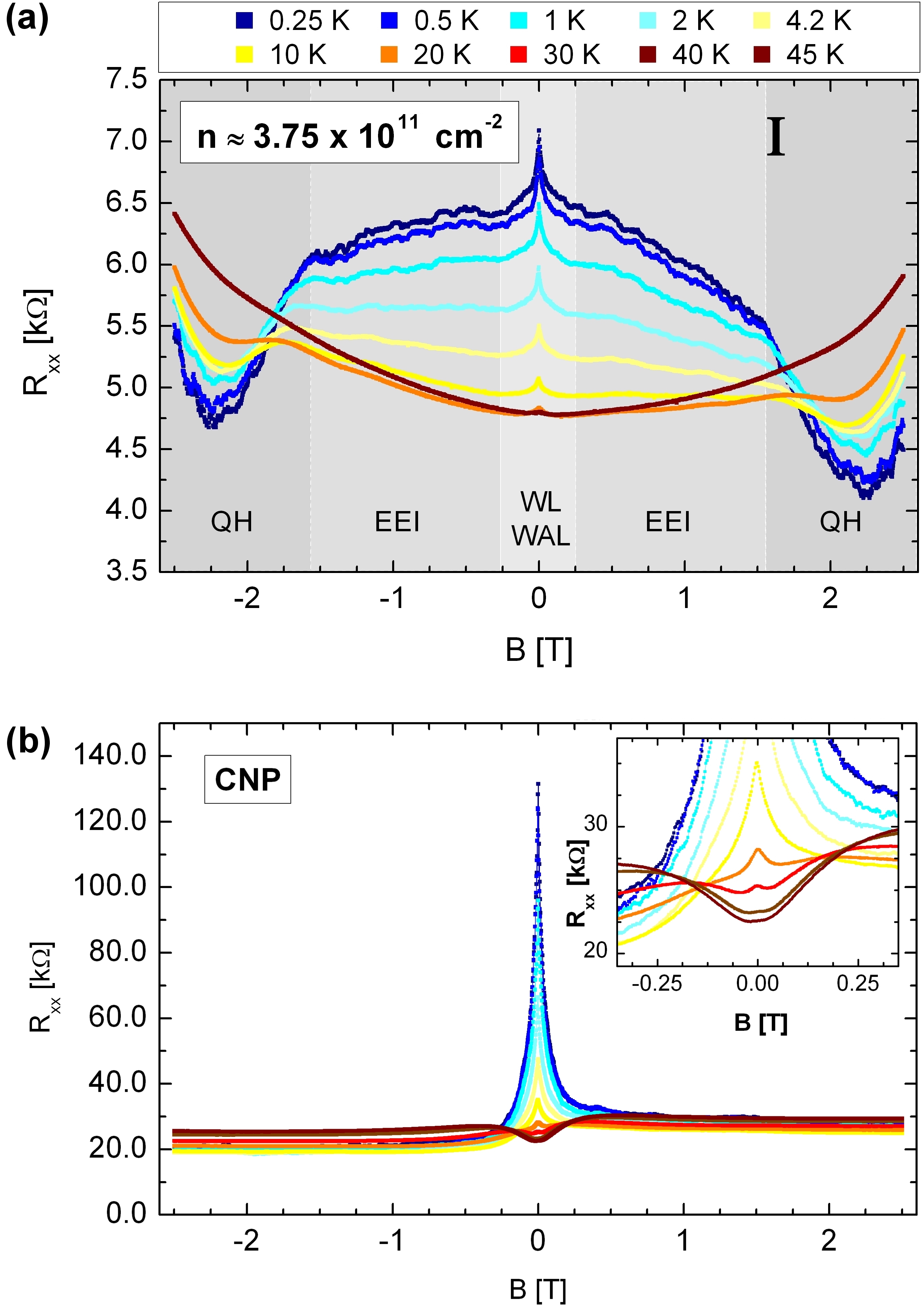}
\caption{(Color online) Longitudinal magnetoresistance at different temperatures measured for (a) point $I$ ($V_{TG}=0$~V) and (b) at the CNP ($V_{TG}=-27$~V). The range of magnetic field in which the different contributions to the magnetoresistance are dominant is highlighted in (a).}
\label{fig3}
\end{figure}

The qualitative features of quantum interference and EEI are illustrated by means of Fig.~\ref{fig3} where we plot the magnetoresistance measured for $V_{TG}=0$~V and for $V_{TG}=-27$~V at different temperatures. In Fig.~\ref{fig3}(a), a positive magnetoresistance peak, centered at $B=0$~T and extending in a narrow range, $\left|B\right|<0.3$ T, is clearly visible. Its amplitude decreases with temperature and reaches almost complete suppression at $T=45$~K. At larger magnetic fields, the magnetoresistance is dominated by EEI, which gives a broad parabolic background,\cite{JouaultPRB2011,JobstPRL2012} discussed in detail in Sec. \ref{secEEI}, and an inversion of concavity when the temperature is increased. Unlike interference corrections (both WL and WAL) that are quickly suppressed as $B$ is increased, the EEI correction survives up to higher magnetic fields and is limited only by the onset of Landau quantization. In the dataset shown in Fig.~\ref{fig3}(a), for low temperatures a kink is observed at $\left|B\right|\approx 1.5$~T, where the first Shubnikov$-$de Haas (SdH) minima start to develop.

The dataset shown in Fig.~\ref{fig3}(b) was measured at a top-gate voltage $V_{TG}=-27$~V. At this bias the Fermi level is approximately located at the CNP (see Figs.~\ref{fig1}(b) and \ref{fig2}) and the charge density is at its minimum. According to Ref.~\onlinecite{TikhonenkoPRL2009}, the most favorable condition to observe WAL is when the carrier density is decreased and the temperature increased. In fact, Fig.~\ref{fig3}(b) shows that the low field magnetoresistance peak evolves into a dip as the temperature is increased, consistent with the passage from WL to WAL interference regimes.\cite{McCannPRL2006} Close to the CNP, the difficulty in estimating the carrier density prevented the application of the analysis reported in the following sections. As a result, for this gate voltage, a contribution of EEI to the evolution of negative to positive magnetoresistance cannot be completely excluded. Another consequence of the lower charge density occurring in Fig.~\ref{fig3}(b) with respect to Fig.~\ref{fig3}(a) is the widening of the magnetic field range in which QH effects dominate, which greatly narrows the field window available for the analysis of quantum interference and EEI.

\subsection{\label{secEEI} Electron Electron Interaction}

EEIs can induce an appreciable contribution to the resistance of 2D materials that extends in the range of non-quantizing magnetic fields beyond the narrow window in which localization effects are observed. For all studied temperatures and carrier densities here, the thermal length $\ell_{th} = \hbar v_{F} / k_{B} T$ in our samples was much larger than the momentum relaxation length $\ell_{0} = v_{F} \tau_{0}$, i.e., $k_{B} T \tau_{0} / \hbar \ll 1$, where $v_{F} = 1.1 \times 10^{6}$~m/s is the quasiparticle Fermi velocity and $\tau_{0}$ the momentum relaxation time, obtained from the zero-field classic (Drude) resistance $R_{0}$,\cite{Drude} which yielded $\tau_{0}\approx 0.01-0.02$~ps.  In this diffusive limit, EEIs are known to cause a correction to the longitudinal conductivity which is logarithmic in $T$ and independent of magnetic field.\cite{AAEEI} Upon tensor inversion this leads to a resistivity correction parabolic in $B$~\cite{AAEEI,Kozikov2010,JobstPRL2012} (and logarithmic in $T$):
\begin{equation}
  \frac{\Delta R_{xx}}{R^{2}_{0}}\approx \left[\left(\omega_{c}\tau_{0}\right)^{2}-1\right]\frac{e^{2}}{2\pi^{2}\hbar}\left[K_{ee} ln\left(\frac{k_{B}T\tau_{0}}{\hbar}\right)\right],
	\label{eqEEI}
\end{equation}
where $\Delta R_{xx} = R_{xx} - R_{0}$. The charge carrier density $n$ enters the EEI correction through the cyclotron frequency $\omega_{c} = \left( v_{F} e / \hbar \sqrt{\pi n}\right)B$, which also includes the $B$--dependence. The  dimensionless quantity $K_{ee}$ is a measure of the strength of the interaction  and depends on several instrinsic and extrinsic parameters such as the spin and valley degeneracy and the dielectric environment of the two-dimensional conductor.

Following Ref.~\onlinecite{JobstPRL2012}, the magnetoresistance curves were first normalized by calculating $(R_{xx}-R_{0})/R^{2}_{0}$, where the value of $R_{0}$ was recursively adjusted to find a best fit to the data. The fits to Eq.~(\ref{eqEEI}) were performed by grouping the dependence on temperature and on the EEI coupling term into the quantity $A=K_{ee} ln\left(k_{B}T\tau_{0}/\hbar\right)$ (i.e., essentially the curvature of the parabolic magnetoresistance) which was then used as fitting parameter. $n$ was set to the experimental values obtained in Fig.~\ref{fig2}. 

\begin{figure}[tbp]
\includegraphics[width=\columnwidth]{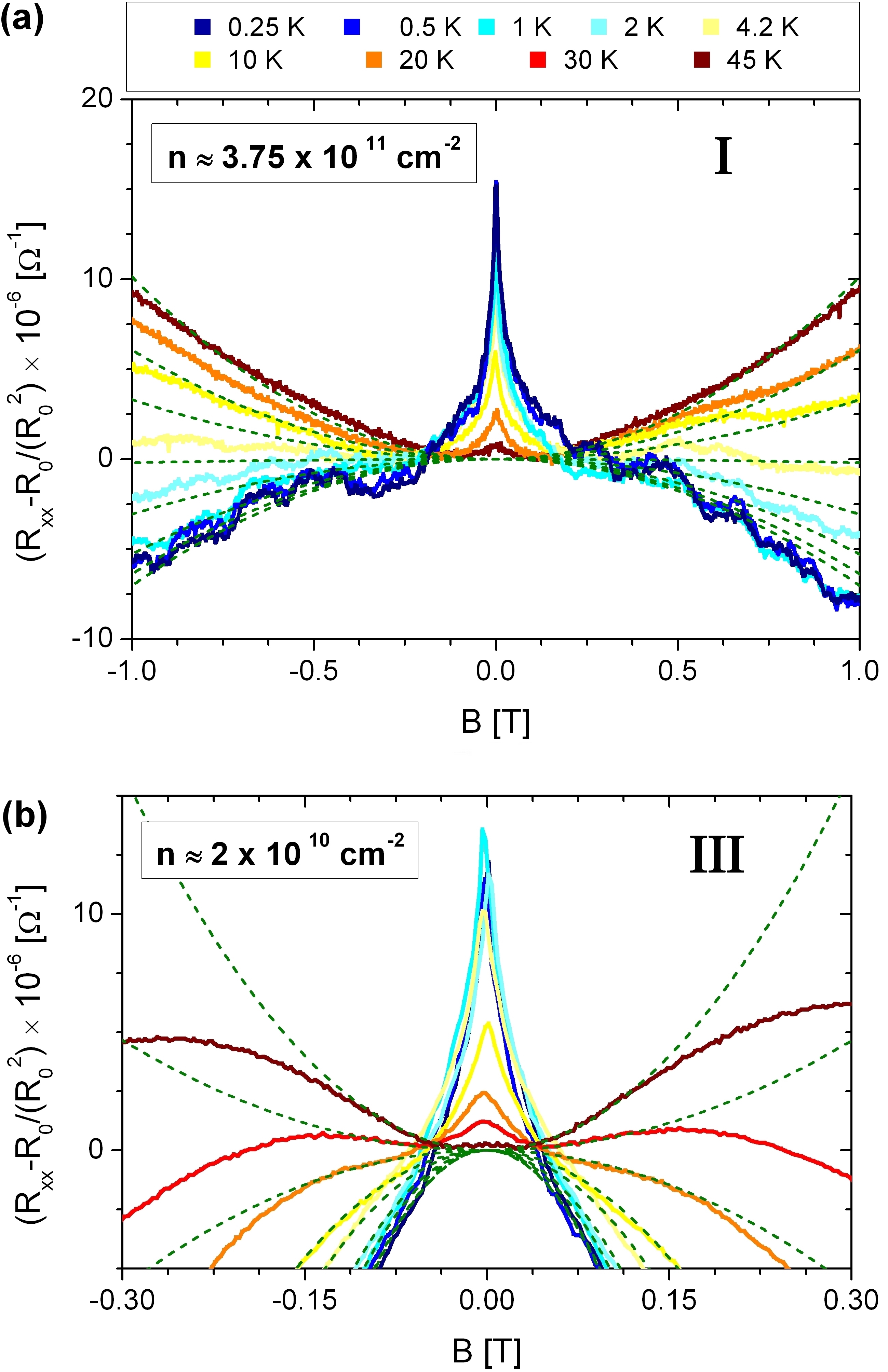}
\caption{(Color online) Extraction of the EEI contribution to the magnetoresistance for density points (a) I and (b) III. For all temperatures, the magnetoresistance is normalized to $(R_{xx}-R_{0})/R_{0}^{2}$, and then fit parabolas according to Eq.~(\ref{eqEEI}). Note the different $B$--range displayed in (a) and (b).}
\label{fig4}
\end{figure}

Figure~\ref{fig4} shows the normalized curves and the results of the best fits for the density points $I$ and $III$ defined in the inset of Fig.~\ref{fig2}. Our data are well described by the EEI correction given by Eq.~(\ref{eqEEI}) and show a quadratic $B$--dependence and an increasing amplitude both in the low and high temperature limits. 

The limits of the fitting procedure were set by the occurrence of the QH effect. As the carrier density was lowered, the QH correction to magnetoresistance became more important because the two symmetric SdH minima approached the $B=0$~T limit. This resulted in the narrowing of the $B$--range available to fit our data, and is clearly visible by comparing Figs.~\ref{fig4}(a) and (b). Since the range $\Delta B$ between two symmetric SdH minima is proportional to $\Delta n$, this effect is rather severe, and prevented the extraction of the EEI correction for some temperatures of dataset $IV$, for which the parabolic background merged with the WL peak.

\begin{figure}[tbp]
\includegraphics[width=\columnwidth]{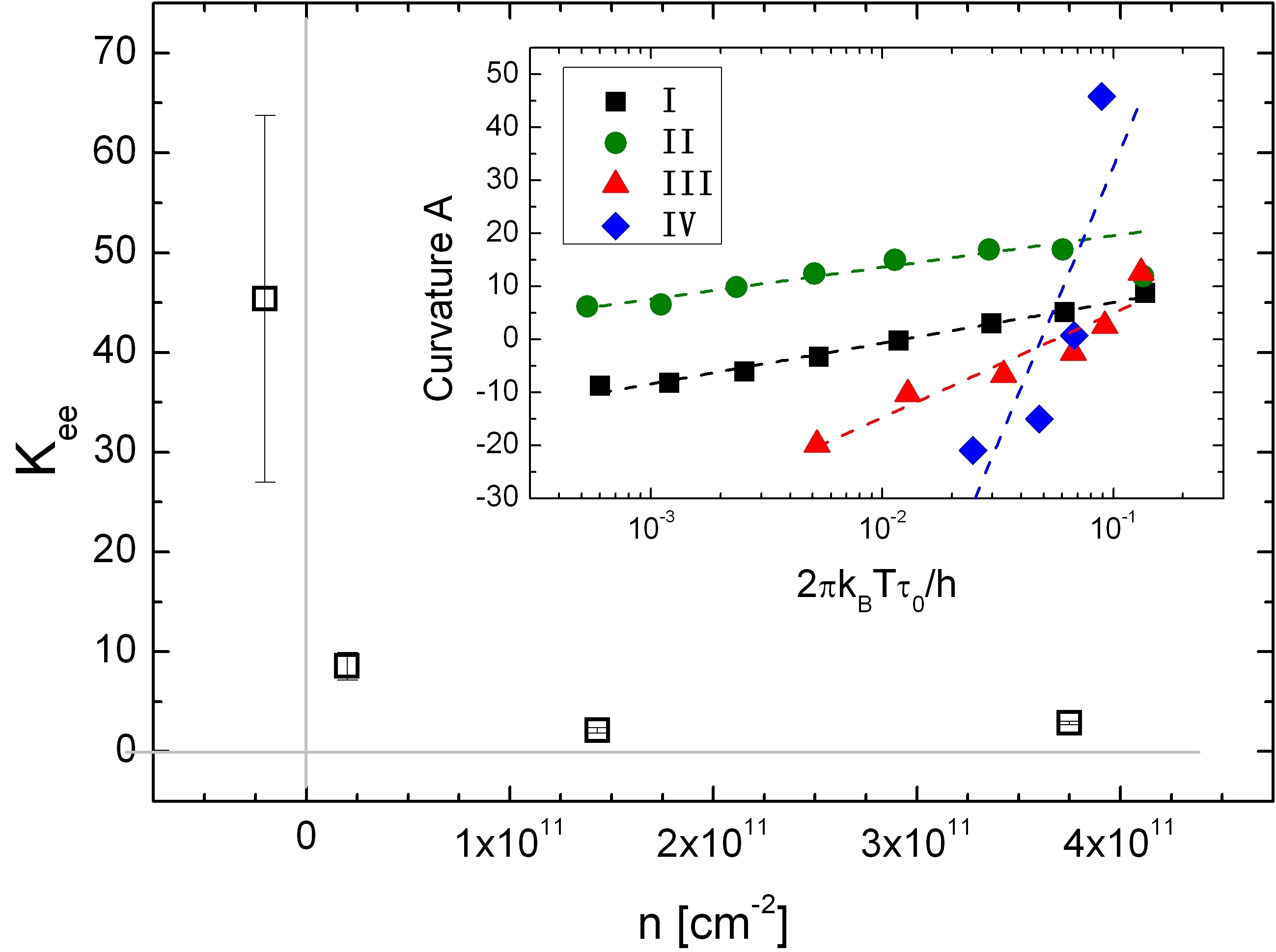}
\caption{(Color online) Dependence of the interaction parameter $K_{ee}$ on carrier density $n$, obtained from the linear fits (dashed lines) of the curvatures $A$ shown in the inset. The error bars are the standard deviations of the fits. In the electron region $(n>0)$, the error bars are smaller than the square symbols.}
\label{fig5}
\end{figure}

An insight into the EEI is obtained by considering the plot of the fitted curvature $A$ as a function of $k_{B}T\tau_{0}/\hbar$ shown in the inset of Fig.~\ref{fig5}. By using a logarithmic scale, we confirm that the data follow the logarithmic behaviour predicted by Eq.~(\ref{eqEEI}) in a broad temperature range spanning more than two orders of magnitude. This is in agreement with what was found in epitaxial and quasi--freestanding graphene,\cite{JobstPRL2012,Lara-AvilaPRL2011} and further indicates that electrons in disordered graphene behave as a Fermi liquid. From Fig.~\ref{fig5}, we first note that the slope for each density, and thus the interaction parameter $K_{ee}$, remains constant in the whole temperature range investigated.

The second information is provided by the dependence of the interaction parameter $K_{ee}$ on the carrier density $n$ obtained from the linear fits of $A$, and shown in the main graph of Fig.~\ref{fig5}. A strong variation of $K_{ee}$ is visible, with an overall decreasing trend as $n$ is increased. This variation is much larger than the experimental error for most of the data. For density point IV, the increased scattering of the $A$ values and the smaller number of points, both arising from the difficulty in fitting the EEI parabola at low carrier density, resulted in a larger error bar. This uncertainty, however, does not affect the observation that $K_{ee}$ is strongly sensitive to changes in carrier density.

This behaviour of $K_{ee}$ is surprising, since the interaction parameter is expected to follow the relation $K_{ee}=1+c(1-\log(1+F^\sigma_0)/F^\sigma_0)$,\cite{Kozikov2010} where $F^\sigma_0$ is the Fermi-liquid constant, and $c$ is linked to the number of multiplets participating in the electron--electron scattering. Peculiar functional forms are expected for $K_{ee}$ in materials such as bilayer graphene, Si(100), and GaAs-based 2DEGs (with $K_{ee}$ generally increasing for decreasing $n$), but $F^\sigma_0$ is predicted to be independent of $n$ in monolayer graphene.\cite{SarmaRMP2011} On the other hand, our results demonstrate clear $n$-dependent EEI signatures but the resulting $K_{ee}$ parameter displays an evolution in the low-$n$ regime which is not straightforward to fit within the existing theory and deserves more experimental and theoretical investigation. Possible reasons for the behavior observed here might be linked to the complex dielectric environment of our gated graphene on SiC and to charge inhomogeneities, whose effect becomes particularly strong near the CNP.

\subsection{\label{secWL} Quantum Interference}

For the analysis of our data, we refer to the theory of quantum interference in graphene developed in Ref.~\onlinecite{McCannPRL2006}, where the correction to the magnetoresistance is found to be
\begin{multline}
\frac{\Delta R_{xx}}{R^{2}_{0}} = - \frac{e^{2}}{\pi h} \left[F\left(\frac{\tau^{-1}_{B}}{\tau^{-1}_{\varphi}}\right)-F\left(\frac{\tau^{-1}_{B}}{\tau^{-1}_{\varphi}+2\tau^{-1}_{iv}}\right) \right.\\ 
\left. -2F\left(\frac{\tau^{-1}_{B}}{\tau^{-1}_{\varphi}+\tau^{-1}_{*}}\right)\right]
\label{eqWL}
\end{multline}
with $F(z)=ln\left(z\right)+\psi\left(0.5+z^{-1}\right)$, $\psi\left(x\right)$ is the di\-gamma function, $\tau^{-1}_{B}=\frac{4 D e B}{\hbar}$, and $D$ is the diffusion coefficient.

\begin{figure}[tbp]
\includegraphics[width=\columnwidth]{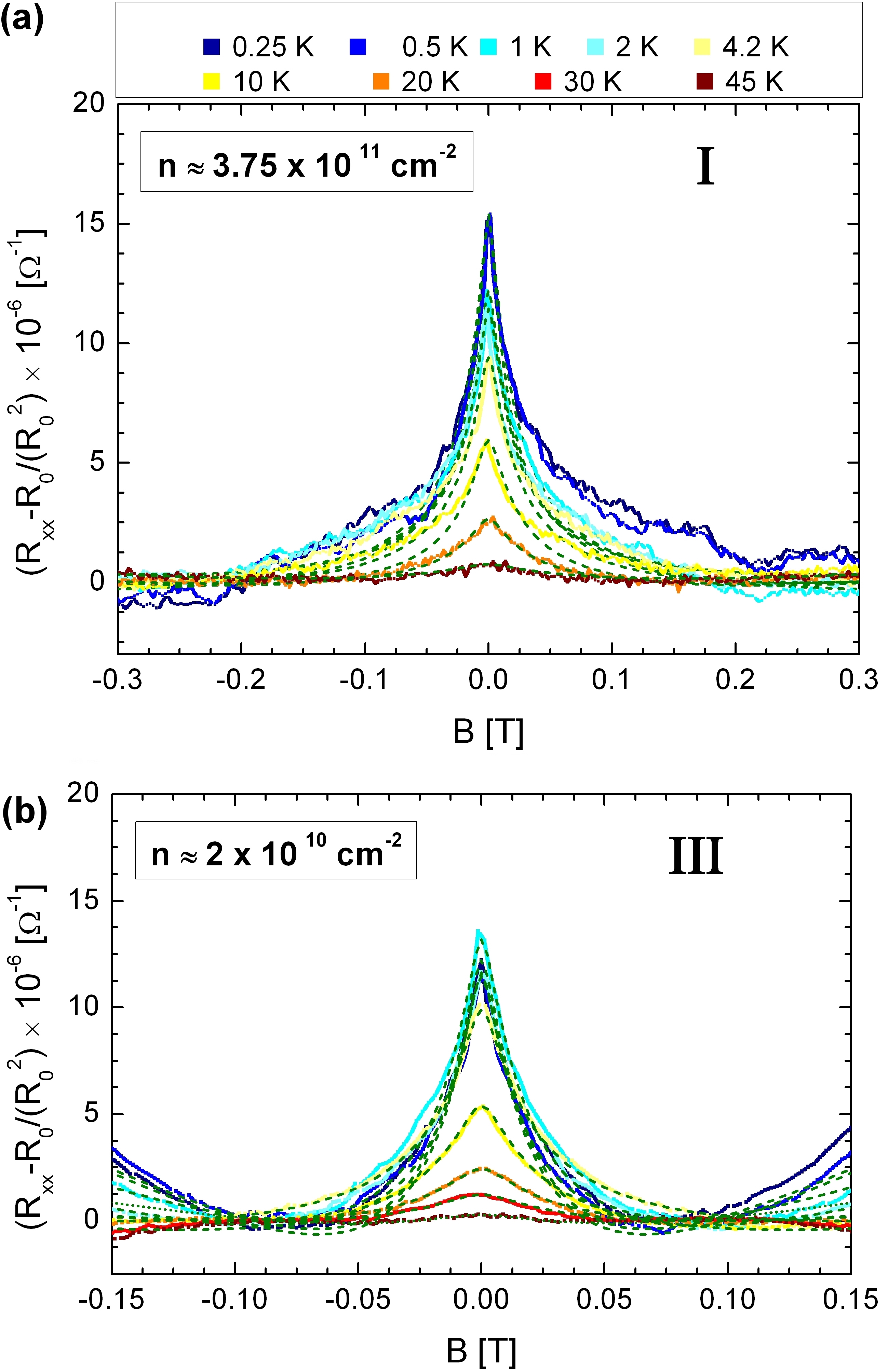}
\caption{(Color online) Fit of the quantum interference contribution to the magnetoresistance, at different temperatures, for the density points (a) I and (b) III. The fits are shown as dashed lines. Note the different $B$--range displayed in (a) and (b).}
\label{fig6}
\end{figure}

Figure~\ref{fig6} shows the normalized magnetoresistance data measured for two values of $n$, after subtracting analytically the EEI correction obtained in Sec.~\ref{secEEI}. The dashed lines are fits to Eq.~(\ref{eqWL}). The fitting procedure, involving three parameters ($\tau_{\varphi}$, $\tau_{iv}$, $\tau_{*}$), is rather delicate, but can be performed by noting that the effect of each scattering time is more evident in distinct field ranges. In particular, while changing $\tau_{\varphi}$ modifies the peak amplitude and width around $B=0$~T, variations in $\tau_{iv}$ determine the width at the base of the peak. Finally, $\tau_{*}$ mainly affects the slope of the magnetoresistance at the sides of the peak.

\begin{figure*}[tbp]
\includegraphics[width=0.9\textwidth]{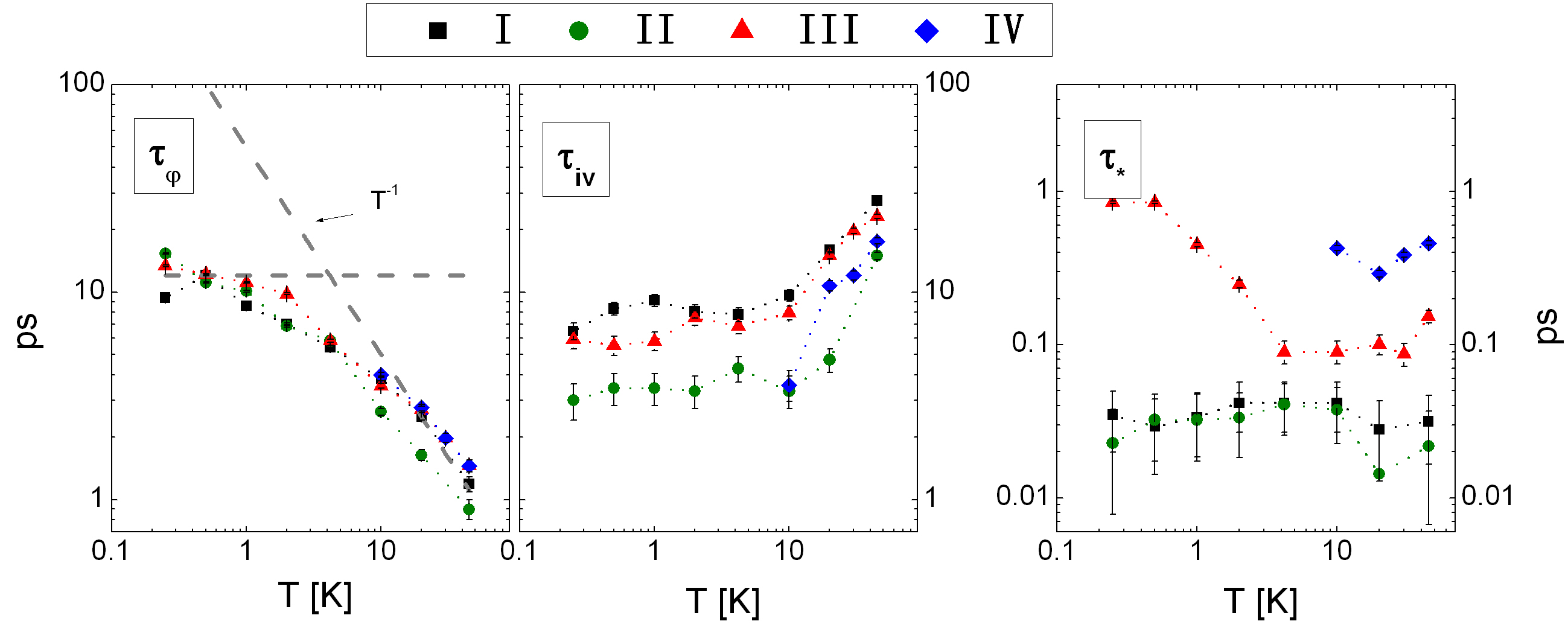} 
\caption{(Color online) Temperature dependence of the scattering times obtained from the fits of the WL corrections. A $\propto T^{-1}$ curve and a constant one, shown as dashed lines in (a), are a guide to the eye. The uncertainty in the scattering times, expressed by the error bars, is estimated as the maximum variation in the values which allow to retain a satisfactory fit.}
\label{fig7}
\end{figure*}

The scattering times obtained from the analysis are shown in Fig.~\ref{fig7} as a function of temperature. The numerical values are of the order of $1 - 10$~ps for the dephasing time $\tau_{\varphi}$, $\sim$10~ps for the intervalley time $\tau_{iv}$, and $0.01 - 1$~ps for the intravalley time $\tau_{*}$. These values are consistent with previously reported data on mechanically exfoliated,\cite{TikhonenkoPRL2008} epitaxial, and CVD graphene devices\cite{BakerPRB2012} measured in the $\sim$10~K temperature range. This indicates that the presence of the SiC substrate, which is expected to have a strong interaction with the graphene layer, actually does not have a dramatic influence on the value of the scattering times. A deeper understanding of the effect of the substrate can be gained by comparing the amplitudes of the elastic scattering times with that of the momentum relaxation time $\tau_{0}$ calculated in Sec.~\ref{secEEI}. In our device, $\tau_{0}\sim 0.01$~ps, so the relation $\tau_{0}\approx \tau_{*} \ll \tau_{iv}$ holds, which indicates that $intra$valley scattering is the main source of disorder in epitaxial graphene. This was already pointed out in previous works on graphene devices on SiC, and it was related to the presence of donors in the buffer layer. \cite{Lara-AvilaPRL2011}

Second, while $\tau_{iv}$ and $\tau_{*}$ show a rather weak temperature dependence in the investigated range, the variation of $\tau_{\varphi}$ is more pronounced, and $\tau_{\varphi}$ decreases with increasing $T$. This behaviour is expected, since high temperature is known to enhance the dephasing of quasiparticles.\cite{TikhonenkoPRL2009} In graphene, inelastic interactions due to electron-electron scattering were found to be the dominant mechanism limiting the coherence of quasiparticles at low temperature.\cite{TikhonenkoPRL2009,JouaultPRB2011,JobstPRL2012} In the diffusive regime, this interaction has a characteristic $T^{-1}$ dependence.\cite{AAEEI} Many papers on different graphene samples reported that a saturation of $\tau_{\varphi}$\cite{BakerPRB2012,JobstPRL2012,JouaultPRB2011,TikhonenkoPRL2008} starts to develop at temperatures below $T\approx10$ K, whose origin is still not well understood. Also our results show a crossover of $\tau_{\varphi}$ between a flat regime, at low temperature, and the $T^{-1}$ dependence predicted for electron-electron scattering, at higher temperature. This behaviour is highlighted in Fig.~\ref{fig7}(a) by two dashed lines, which indicate an approximate crossover temperature of 4~K.

Next, we consider the behaviour of the scattering times on carrier density $n$. The dephasing time $\tau_{\varphi}$ does not display any variation with $n$ in the investigated range, and the different curves almost fall on top of each other. The $inter$valley scattering time $\tau_{iv}$ shows small fluctuations, but no clear dependence on $n$. 
Such scattering of data, whose origin is not clear, was already observed in mechanically exfoliated devices with slightly larger ($\sim 10^{12}$~cm$^{-2}$) carrier densities (see supplementary material of Ref.~\onlinecite{TikhonenkoPRL2008}), where also values of $\tau_{iv}\approx 10$~ps were measured.

On the other hand, $\tau_{*}$ increases appreciably with decreasing density. This is consistent with our observation of more pronounced WAL effects in the magnetoresistance curves measured in proximity to the CNP, where the most favorable conditions\cite{TikhonenkoPRL2009} to observe WAL effects are reached (cf.~Fig.~\ref{fig3}). The stronger variation of $intra$valley scattering, as compared to both $inter$valley scattering and dephasing, appears to be a general property of graphene devices, as highlighted by the comprehensive collection of data reported in Ref. \onlinecite{BakerPRB2012}, where devices fabricated with different methods were compared. In particular, a weakening of $intra$valley scattering is generally observed with decreasing carrier density, which implies that the low-density region must be explored in order to improve the performance of graphene--based devices.

\begin{figure}[tbp]
\includegraphics[width=\columnwidth]{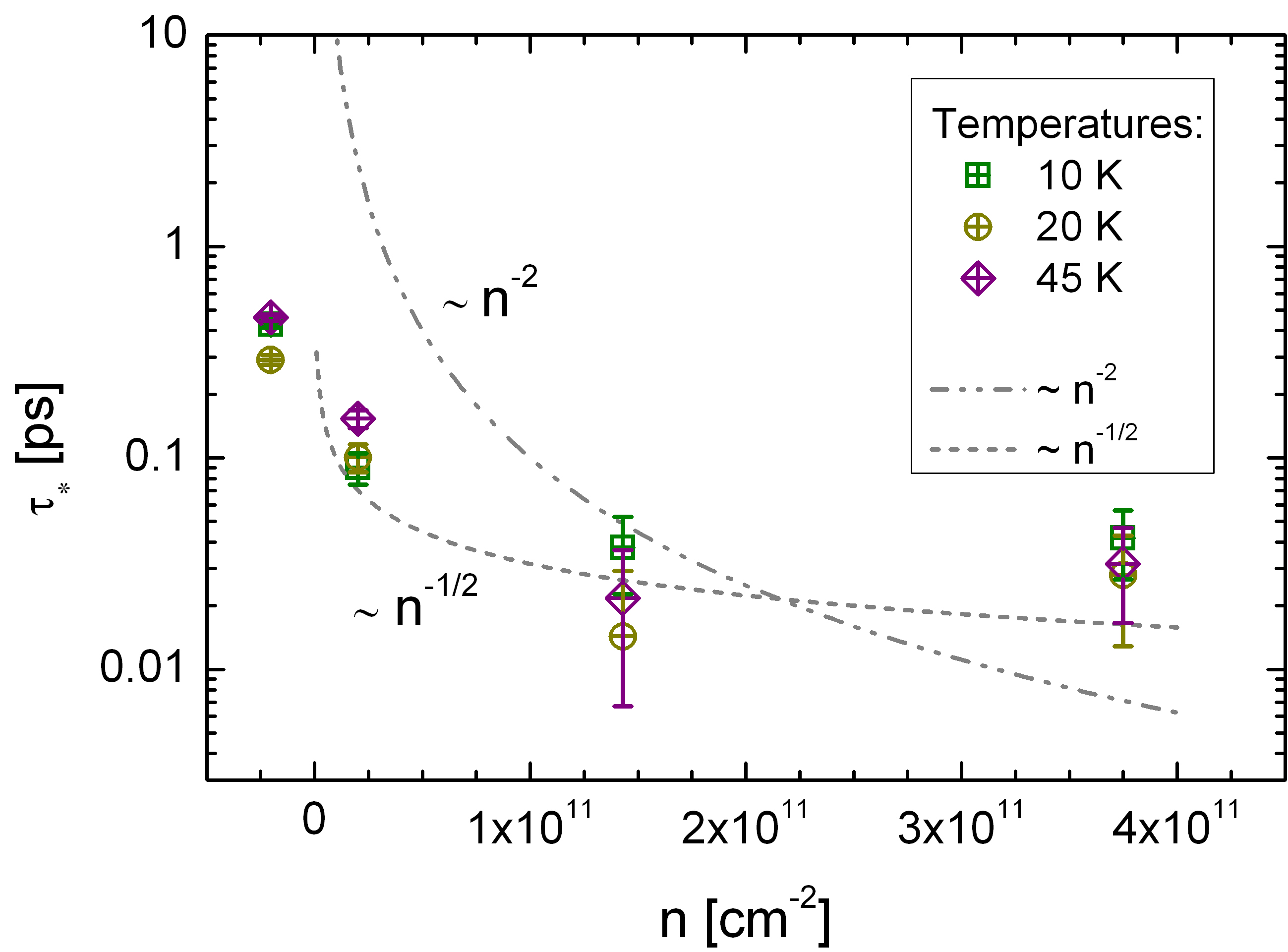}
\caption{(Color online) Dependence of the $intra$valley time $\tau_{*}$ on charge density $n$ (same error bars as in Fig.~\ref{fig7}). The data are compared with the $\propto n^{-2}$ dependence predicted for the scattering time due to trigonal warping. A $\propto n^{-1/2}$ curve is also shown.}
\label{fig8}
\end{figure}

In Fig.~\ref{fig8}, we show $\tau_{*}$ for the three highest temperatures in our range as a function of the charge density $n$. The data show a clear decreasing trend with increasing density, with a variation of one order of magnitude in the investigated density range. A dependence of the $intra$valley scattering on the carrier density is expected through the warping term, since trigonal warping depends on the Fermi energy,\cite{McCannPRL2006} and becomes the stronger the further away from the CNP. On the other hand, other chirality-breaking mechanisms based on short-range defects and device edges are expected to be insensitive to changes in the charge density.\cite{TikhonenkoPRL2008}

To investigate the origin of the behaviour of $\tau_{*}$ shown in Fig.~\ref{fig8}, we refer to the theory developed in Ref.~\onlinecite{McCannPRL2006}, where a functional dependence $\propto n^{-2}$ was found for the warping scattering time. By comparing our data with a $\propto n^{-2}$ curve, shown in the same Figure, we find that $\tau_{*}$ has a weaker dependence. An analogue result was observed in Ref.~\onlinecite{BakerPRB2012} on the chirality$-$breaking scattering length $L_{*}=\sqrt{D\tau_{*}}$, where a dependence $L_{*}\propto n^{-1/4}$ was found. In terms of scattering times, their observation corresponds to a $\tau_{*}\propto n^{-1/2}$ behaviour. From the comparison of our data with a $n^{-1/2}$ curve, shown in Fig.~\ref{fig8}, we confirm the dependence found in Ref.~\onlinecite{BakerPRB2012}. This suggests that chirality-breaking scattering due to trigonal warping, although important, is not the dominant contribution of $intra$valley scattering, but also scattering due to sharp topological defects such as adatoms, vacancies, pentagons or heptagons has to be taken into account.

\section{\label{secConc} Conclusion}

In conclusion, we presented a systematic analysis of the magnetotransport properties in epitaxial graphene grown on the Si-terminated face of SiC, and we extract the two quantum corrections affecting the low-field magnetoresistance -- quantum interference and EEI. The possibility of tuning the charge density by means of a top gate enabled us to control the magnitude of the two quantum contributions, and to investigate the combined effect of density and temperature. 

We successfully describe the main features of EEI in graphene with the current theory for disordered systems. However, we find evidence for an unexpected dependence of the interaction parameter $K_{ee}$ on carrier density, not accounted for by theory.

From fits of the quantum interference correction, we obtain the dependence of the scattering times on carrier density. In particular, we find that while the dephasing and $inter$valley scattering times are almost constant, the $intra$valley scattering time shows a peculiar dependence in the investigated density range, which is different from the one arising from the sole warping term.

Our results stress the role of charge density in determining the properties of both quantum interference and EEI, and the necessity of a further investigation of its impact on the low-field magnetoresistance of graphene. 

\begin{acknowledgments}
The authors would like to thank Fabio Beltram for critical reading of the manuscript. Furthermore, we acknowledge financial support from the Italian Ministry
of Research (MIUR-FIRB project RBID08B3FM).
\end{acknowledgments}

\appendix

\section{\label{APP}}

Macroscopic sample inhomogeneities, such as small gradients in the charge density, geometrical effects, and contact misalignment, are known to introduce artifacts in the measured magnetotransport quantities. All these macroscopic effects result in a dependence of the measured transport quantities on the choice of the particular contact pairs used for the measurement. 

From an experimental point of view, the impact of these effects can be relevant, and complicates the investigation of the transport properties. In particular, different values of longitudinal resistance can be measured at opposite sides of the Hall bar, and a dependence of the amplitude of the SdH oscillations on the polarity of magnetic field is often observed. Most of the studies on these aspects are on 2D semiconductor devices (see for instance Ref.~\onlinecite{PonomarenkoSSComm2004} and reference therein), while little discussion is dedicated to graphene. In graphene, it is common practice\cite{JouaultPRB2011,JobstPRL2012} to perform a symmetrization of the data with respect to magnetic field. This is done by averaging the measured resistance $R_{xx}(B)$ with the resistance $R_{xx}(-B)$ obtained upon inverting the polarity of $B$.

\begin{figure}[tbp]
\includegraphics[width=\columnwidth]{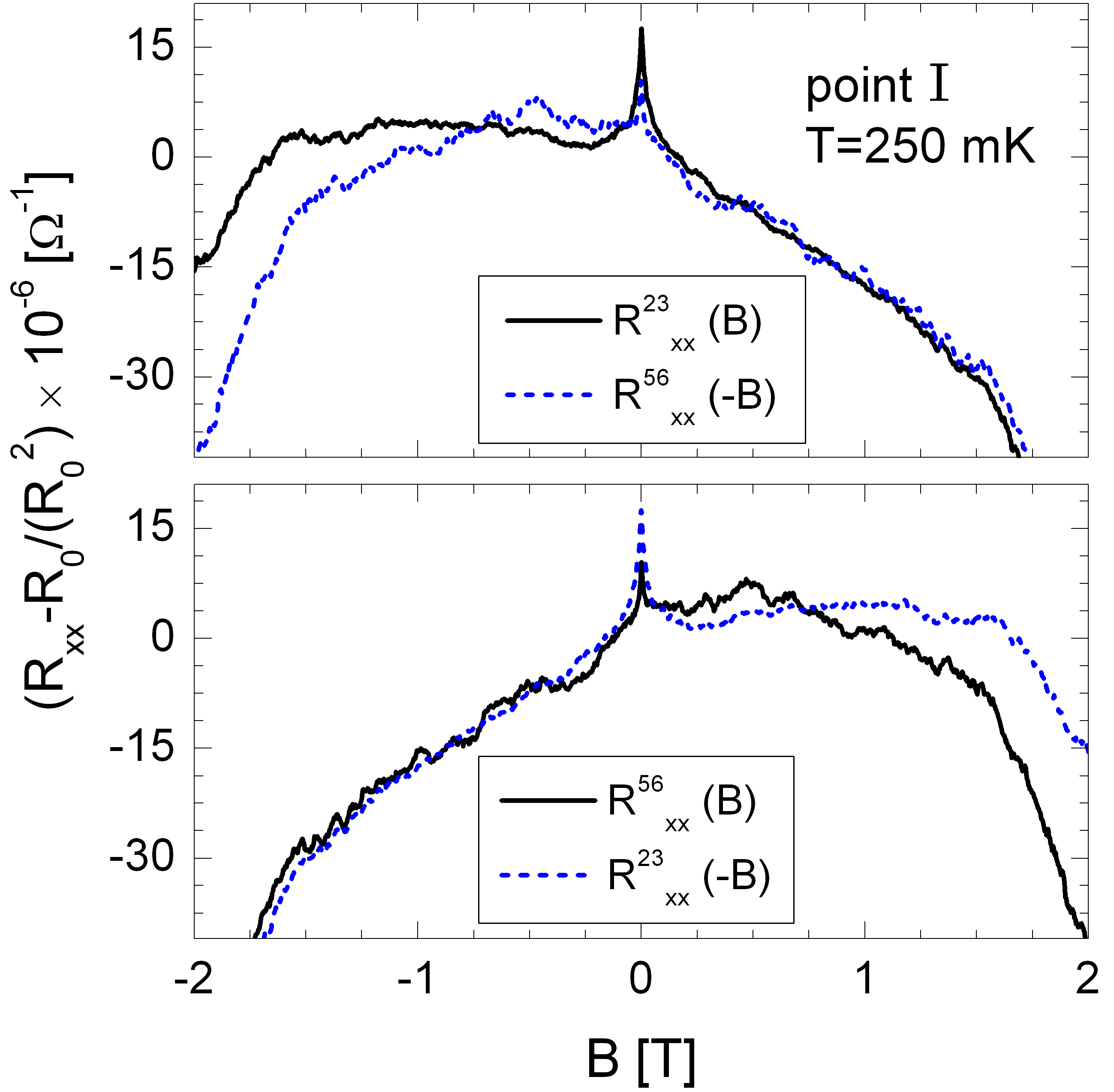}
\caption{(Color online) Comparison between the magnetoresistances $R^{23}_{xx}(B)$ and $R^{56}_{xx}(B)$, shown as solid lines, with $R^{23}_{xx}(-B)$ and $R^{56}_{xx}(-B)$, displayed as dashed line, obtained after the inversion of polarity of magnetic field.}
\label{fig9}
\end{figure}

In this work, we make use of an alternative method to account for the macroscopic inhomogeneities of our device. We concentrate on gradients of the charge density introduced by the fabrication technology used for top-gating the Hall bar, since geometric factors (e.g., contact misalignment or deviations from the rectangular shape) are secondary in our device. Small density gradients along the channel direction of the Hall bar introduce a peculiar $B$--dependence in $R_{xx}$, resulting in a ``tilt" of the magnetoresistance curve, which is accompanied by characteristic symmetries on $B$. For linear density gradients, theory\cite{KarmakarPhysE2004} predicts an anti$-$symmetric behaviour of the magnetoresistance measured at two opposite sides of the Hall bar, and in this case, the best estimate of the longitudinal magnetoresistance $R_{xx}$ is obtained by averaging the two quantities. Referring to Fig.~\ref{fig1} for the meaning of symbols, this results in the symmetry conditions
\begin{flalign}
R^{23}_{xx}(B) = R^{56}_{xx}(-B),\label{eqCond1}\\ 
R^{56}_{xx}(B) = R^{23}_{xx}(-B),\label{eqCond2}
\end{flalign}
and $R_{xx}=\frac{1}{2}(R^{23}_{xx}+R^{56}_{xx})$.

In Fig.~\ref{fig9}, we plot the measured magnetoresistances $R^{23}_{xx}(B)$ and $R^{56}_{xx}(B)$, together with the mirrored quantities $R^{56}_{xx}(-B)$ and $R^{23}_{xx}(-B)$ obtained by inverting the $B$--polarity. The data are displayed as $\Delta R_{xx}/R_{0}^{2}$ to allow for a comparison with Fig.~\ref{fig4} and with Eqs.~(\ref{eqEEI}) and (\ref{eqWL}). The value of $R_{0}$ was calculated as the average of $R_{xx}(B)$ at $B=+0.2$~T and $B=-0.2$~T to avoid the effect of WL. From the figure, we see that $R^{23}_{xx}(B)$ and $R^{56}_{xx}(B)$ fulfil the symmetry conditions in Eqs.~(\ref{eqCond1}) and (\ref{eqCond2}) for most of the magnetic field range, which indicates that a linear component of the charge density gradient is present in our device. The non-perfect overlap of the two pairs of curves causes the residual ``tilt" visible in the magnetoresistance set in Fig.~\ref{fig3}, where the magnetoresistance $R_{xx}$ is shown. We attribute this small residual distortion to higher order terms in the density gradients, as discussed in detail in the following.

\begin{figure}[tbp]
\includegraphics[width=\columnwidth]{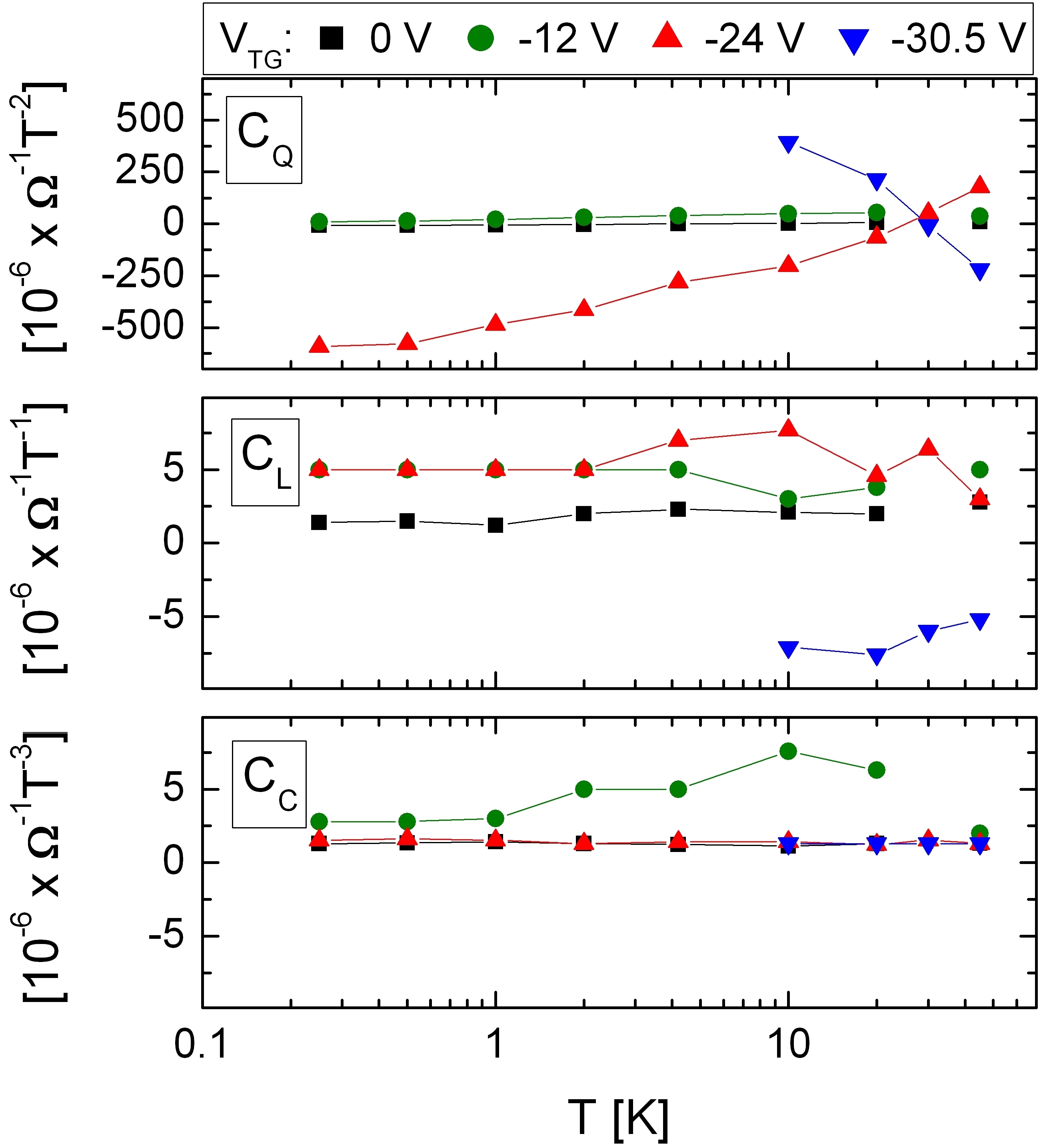}
\caption{(Color online) Coefficients of correction to the magnetoresistance due to non-linear variation of the charge density. Note the different scale of the quadratic term $C_{Q}$ (upper panel) compared to the other terms.}
\label{fig10}
\end{figure}

In general, non-linear density gradients give rise to a more complex problem, and an accurate treatment of it involves the use of higher order terms. In this case, the best estimate $R^{NLin}_{xx}$ is\cite{KarmakarPhysE2004}
\begin{equation}
\frac{R^{23}_{xx}+R^{56}_{xx}}{2}=R^{NLin}_{xx}F(\omega_{c}\tau),
\label{eqGeom1}
\end{equation}
where $F(\omega_{c}\tau)$ is a polynomial function of the cyclotron frequency $\omega_{c}$. As a result, the magnetoresistance correction due to charge density inhomogeneity can be written in the general form, up to third order,
\begin{equation}
\frac{\Delta R_{xx}}{R^{2}_{0}}=C_{Q} B^{2}-C_{L} B - C_{C} B^{3},
\label{eqGeom2}
\end{equation}
where the coefficients $C_{i}$ (i=Q,L,C) are used heuristically to account for both the linear and non-linear effect of the density gradient. 

To correct for non-linear terms in our data, we first calculate $R_{xx}=\frac{1}{2}(R^{23}_{xx}+R^{56}_{xx})$, and then fit the resulting curves to Eq.~(\ref{eqGeom2}) to obtain the parameters $C_{i}$ (i=L, Q, C). Figure~\ref{fig10} shows the values of the parameters $C_{Q}$, $C_{L}$ and $C_{C}$ obtained from the fits for all  datasets used in this work. Two striking features are evident from the Figure: first, the values of $C_{L}$ and $C_{C}$ are always much smaller than the quadratic coefficient $C_{Q}$; second, while $C_{L}$ and $C_{C}$ maintain their sign in the whole temperature range and for all carrier densities, $C_{Q}$ changes sign at high temperatures. These two aspects indicate that the coefficient $C_{Q}$ obtained from the fit have to be identified, up to some constants, with the curvature $A$ used for the analysis of Sec.~\ref{secEEI}. A quadratic correction to the magnetoresistance due to non-linear terms, although present, is instead rather small, and can be neglected. As a consequence, the effect of charge density inhomogeneity was removed from our data first by averaging the magnetoresistance curves measured at opposite sides of the device, and then by subtracting analytically the linear and cubic contributions.

\bibliography{PRB_GrapheneBiblio}

\begin{thebibliography}{34}
\expandafter\ifx\csname natexlab\endcsname\relax\def\natexlab#1{#1}\fi
\expandafter\ifx\csname bibnamefont\endcsname\relax
  \def\bibnamefont#1{#1}\fi
\expandafter\ifx\csname bibfnamefont\endcsname\relax
  \def\bibfnamefont#1{#1}\fi
\expandafter\ifx\csname citenamefont\endcsname\relax
  \def\citenamefont#1{#1}\fi
\expandafter\ifx\csname url\endcsname\relax
  \def\url#1{\texttt{#1}}\fi
\expandafter\ifx\csname urlprefix\endcsname\relax\def\urlprefix{URL }\fi
\providecommand{\bibinfo}[2]{#2}
\providecommand{\eprint}[2][]{\url{#2}}

\bibitem[{\citenamefont{Novoselov}(2011)}]{NovoselovRMP2011}
\bibinfo{author}{\bibfnamefont{K.~S.} \bibnamefont{Novoselov}},
  \bibinfo{journal}{Rev. Mod. Phys.} \textbf{\bibinfo{volume}{83}},
  \bibinfo{pages}{837} (\bibinfo{year}{2011}).

\bibitem[{\citenamefont{Lin et~al.}(2010)\citenamefont{Lin, Dimitrakopoulos,
  Jenkins, Farmer, Chiu, Grill, and Avouris}}]{LinScience2010}
\bibinfo{author}{\bibfnamefont{Y.-M.} \bibnamefont{Lin}},
  \bibinfo{author}{\bibfnamefont{C.}~\bibnamefont{Dimitrakopoulos}},
  \bibinfo{author}{\bibfnamefont{K.~A.} \bibnamefont{Jenkins}},
  \bibinfo{author}{\bibfnamefont{D.~B.} \bibnamefont{Farmer}},
  \bibinfo{author}{\bibfnamefont{H.-Y.} \bibnamefont{Chiu}},
  \bibinfo{author}{\bibfnamefont{A.}~\bibnamefont{Grill}}, \bibnamefont{and}
  \bibinfo{author}{\bibfnamefont{P.}~\bibnamefont{Avouris}},
  \bibinfo{journal}{Science} \textbf{\bibinfo{volume}{327}},
  \bibinfo{pages}{662} (\bibinfo{year}{2010}).

\bibitem[{\citenamefont{Bourzac}(2012)}]{BourzacNat2012}
\bibinfo{author}{\bibfnamefont{K.}~\bibnamefont{Bourzac}},
  \bibinfo{journal}{Nature} \textbf{\bibinfo{volume}{483}},
  \bibinfo{pages}{S34} (\bibinfo{year}{2012}).

\bibitem[{\citenamefont{Tzalenchuk et~al.}(2010)\citenamefont{Tzalenchuk,
  Lara-Avila, Kalaboukhov, Paolillo, Syvajarvi, Yakimova, Kazakova, Janssen,
  Fal'ko, and Kubatkin}}]{TzalenchukNatNano2010}
\bibinfo{author}{\bibfnamefont{A.}~\bibnamefont{Tzalenchuk}},
  \bibinfo{author}{\bibfnamefont{S.}~\bibnamefont{Lara-Avila}},
  \bibinfo{author}{\bibfnamefont{A.}~\bibnamefont{Kalaboukhov}},
  \bibinfo{author}{\bibfnamefont{S.}~\bibnamefont{Paolillo}},
  \bibinfo{author}{\bibfnamefont{M.}~\bibnamefont{Syvajarvi}},
  \bibinfo{author}{\bibfnamefont{R.}~\bibnamefont{Yakimova}},
  \bibinfo{author}{\bibfnamefont{O.}~\bibnamefont{Kazakova}},
  \bibinfo{author}{\bibfnamefont{T.~J. B.~M.} \bibnamefont{Janssen}},
  \bibinfo{author}{\bibfnamefont{V.}~\bibnamefont{Fal'ko}}, \bibnamefont{and}
  \bibinfo{author}{\bibfnamefont{S.}~\bibnamefont{Kubatkin}},
  \bibinfo{journal}{Nature Nanotech} \textbf{\bibinfo{volume}{5}},
  \bibinfo{pages}{186} (\bibinfo{year}{2010}).

\bibitem[{\citenamefont{Altshuler et~al.}(1980)\citenamefont{Altshuler,
  Khmel'nitzkii, Larkin, and Lee}}]{AltshulerPRB1980}
\bibinfo{author}{\bibfnamefont{B.~L.} \bibnamefont{Altshuler}},
  \bibinfo{author}{\bibfnamefont{D.}~\bibnamefont{Khmel'nitzkii}},
  \bibinfo{author}{\bibfnamefont{A.~I.} \bibnamefont{Larkin}},
  \bibnamefont{and} \bibinfo{author}{\bibfnamefont{P.~A.} \bibnamefont{Lee}},
  \bibinfo{journal}{Phys. Rev. B} \textbf{\bibinfo{volume}{22}},
  \bibinfo{pages}{5142} (\bibinfo{year}{1980}).

\bibitem[{\citenamefont{Abrahams et~al.}(1979)\citenamefont{Abrahams, Anderson,
  Licciardello, and Ramakrishnan}}]{PhysRevLett.42.673}
\bibinfo{author}{\bibfnamefont{E.}~\bibnamefont{Abrahams}},
  \bibinfo{author}{\bibfnamefont{P.~W.} \bibnamefont{Anderson}},
  \bibinfo{author}{\bibfnamefont{D.~C.} \bibnamefont{Licciardello}},
  \bibnamefont{and} \bibinfo{author}{\bibfnamefont{T.~V.}
  \bibnamefont{Ramakrishnan}}, \bibinfo{journal}{Phys. Rev. Lett.}
  \textbf{\bibinfo{volume}{42}}, \bibinfo{pages}{673} (\bibinfo{year}{1979}).

\bibitem[{\citenamefont{Hikami et~al.}(1980)\citenamefont{Hikami, Larkin, and
  Nagaoka}}]{HikamiPTP1980}
\bibinfo{author}{\bibfnamefont{S.}~\bibnamefont{Hikami}},
  \bibinfo{author}{\bibfnamefont{A.~I.} \bibnamefont{Larkin}},
  \bibnamefont{and} \bibinfo{author}{\bibfnamefont{Y.}~\bibnamefont{Nagaoka}},
  \bibinfo{journal}{Prog. Theor. Phys.} \textbf{\bibinfo{volume}{63}},
  \bibinfo{pages}{707} (\bibinfo{year}{1980}).

\bibitem[{\citenamefont{Bergmann}(1983)}]{BergmannPRB1983}
\bibinfo{author}{\bibfnamefont{G.}~\bibnamefont{Bergmann}},
  \bibinfo{journal}{Phys. Rev. B} \textbf{\bibinfo{volume}{28}},
  \bibinfo{pages}{2914} (\bibinfo{year}{1983}).

\bibitem[{\citenamefont{Bergmann}(1984)}]{BergmannPR1984}
\bibinfo{author}{\bibfnamefont{G.}~\bibnamefont{Bergmann}},
  \bibinfo{journal}{Phys. Rep.} \textbf{\bibinfo{volume}{107}},
  \bibinfo{pages}{1 } (\bibinfo{year}{1984}).

\bibitem[{\citenamefont{Novoselov et~al.}(2005)\citenamefont{Novoselov, Geim,
  Morozov, Jiang, Katsnelson, Grigorieva, Dubonos, and
  Firsov}}]{NovoselovNat2005}
\bibinfo{author}{\bibfnamefont{K.~S.} \bibnamefont{Novoselov}},
  \bibinfo{author}{\bibfnamefont{A.~K.} \bibnamefont{Geim}},
  \bibinfo{author}{\bibfnamefont{S.~V.} \bibnamefont{Morozov}},
  \bibinfo{author}{\bibfnamefont{D.}~\bibnamefont{Jiang}},
  \bibinfo{author}{\bibfnamefont{M.~I.} \bibnamefont{Katsnelson}},
  \bibinfo{author}{\bibfnamefont{I.~V.} \bibnamefont{Grigorieva}},
  \bibinfo{author}{\bibfnamefont{S.~V.} \bibnamefont{Dubonos}},
  \bibnamefont{and} \bibinfo{author}{\bibfnamefont{A.~A.}
  \bibnamefont{Firsov}}, \bibinfo{journal}{Nature}
  \textbf{\bibinfo{volume}{438}}, \bibinfo{pages}{197} (\bibinfo{year}{2005}).

\bibitem[{\citenamefont{Huertas-Hernando
  et~al.}(2006)\citenamefont{Huertas-Hernando, Guinea, and
  Brataas}}]{HuertasPRB2006}
\bibinfo{author}{\bibfnamefont{D.}~\bibnamefont{Huertas-Hernando}},
  \bibinfo{author}{\bibfnamefont{F.}~\bibnamefont{Guinea}}, \bibnamefont{and}
  \bibinfo{author}{\bibfnamefont{A.}~\bibnamefont{Brataas}},
  \bibinfo{journal}{Phys. Rev. B} \textbf{\bibinfo{volume}{74}},
  \bibinfo{pages}{155426} (\bibinfo{year}{2006}).

\bibitem[{\citenamefont{McCann et~al.}(2006)\citenamefont{McCann, Kechedzhi,
  Fal'ko, Suzuura, Ando, and Altshuler}}]{McCannPRL2006}
\bibinfo{author}{\bibfnamefont{E.}~\bibnamefont{McCann}},
  \bibinfo{author}{\bibfnamefont{K.}~\bibnamefont{Kechedzhi}},
  \bibinfo{author}{\bibfnamefont{V.~I.} \bibnamefont{Fal'ko}},
  \bibinfo{author}{\bibfnamefont{H.}~\bibnamefont{Suzuura}},
  \bibinfo{author}{\bibfnamefont{T.}~\bibnamefont{Ando}}, \bibnamefont{and}
  \bibinfo{author}{\bibfnamefont{B.~L.} \bibnamefont{Altshuler}},
  \bibinfo{journal}{Phys. Rev. Lett.} \textbf{\bibinfo{volume}{97}},
  \bibinfo{pages}{146805} (\bibinfo{year}{2006}).

\bibitem[{\citenamefont{Morpurgo and Guinea}(2006)}]{MorpurgoPRL2006}
\bibinfo{author}{\bibfnamefont{A.~F.} \bibnamefont{Morpurgo}} \bibnamefont{and}
  \bibinfo{author}{\bibfnamefont{F.}~\bibnamefont{Guinea}},
  \bibinfo{journal}{Phys. Rev. Lett.} \textbf{\bibinfo{volume}{97}},
  \bibinfo{pages}{196804} (\bibinfo{year}{2006}).

\bibitem[{\citenamefont{Fal'ko et~al.}(2007)\citenamefont{Fal'ko, Kechedzhi,
  McCann, Altshuler, Suzuura, and Ando}}]{FalkoSSC2007}
\bibinfo{author}{\bibfnamefont{V.~I.} \bibnamefont{Fal'ko}},
  \bibinfo{author}{\bibfnamefont{K.}~\bibnamefont{Kechedzhi}},
  \bibinfo{author}{\bibfnamefont{E.}~\bibnamefont{McCann}},
  \bibinfo{author}{\bibfnamefont{B.}~\bibnamefont{Altshuler}},
  \bibinfo{author}{\bibfnamefont{H.}~\bibnamefont{Suzuura}}, \bibnamefont{and}
  \bibinfo{author}{\bibfnamefont{T.}~\bibnamefont{Ando}},
  \bibinfo{journal}{Solid State Commun.} \textbf{\bibinfo{volume}{143}},
  \bibinfo{pages}{33 } (\bibinfo{year}{2007}).

\bibitem[{\citenamefont{Altshuler and Aronov}(1985)}]{AAEEI}
\bibinfo{author}{\bibfnamefont{B.~L.} \bibnamefont{Altshuler}}
  \bibnamefont{and} \bibinfo{author}{\bibfnamefont{A.~G.}
  \bibnamefont{Aronov}}, \emph{\bibinfo{title}{Electron-Electron Interaction in
  Disordered Conductors}} (\bibinfo{publisher}{North Holland},
  \bibinfo{address}{Amsterdam}, \bibinfo{year}{1985}),
  vol.~\bibinfo{volume}{10} of \emph{\bibinfo{series}{Modern Problems in
  Condensed Matter Sciences}}, chap.~\bibinfo{chapter}{1}, pp.
  \bibinfo{pages}{1 -- 155}.

\bibitem[{\citenamefont{Li et~al.}(2003)\citenamefont{Li, Proskuryakov,
  Savchenko, Linfield, and Ritchie}}]{LiPRL2003}
\bibinfo{author}{\bibfnamefont{L.}~\bibnamefont{Li}},
  \bibinfo{author}{\bibfnamefont{Y.~Y.} \bibnamefont{Proskuryakov}},
  \bibinfo{author}{\bibfnamefont{A.~K.} \bibnamefont{Savchenko}},
  \bibinfo{author}{\bibfnamefont{E.~H.} \bibnamefont{Linfield}},
  \bibnamefont{and} \bibinfo{author}{\bibfnamefont{D.~A.}
  \bibnamefont{Ritchie}}, \bibinfo{journal}{Phys. Rev. Lett.}
  \textbf{\bibinfo{volume}{90}}, \bibinfo{pages}{076802}
  (\bibinfo{year}{2003}).

\bibitem[{\citenamefont{Gornyi and Mirlin}(2004)}]{GornyiPRB2004}
\bibinfo{author}{\bibfnamefont{I.~V.} \bibnamefont{Gornyi}} \bibnamefont{and}
  \bibinfo{author}{\bibfnamefont{A.~D.} \bibnamefont{Mirlin}},
  \bibinfo{journal}{Phys. Rev. B} \textbf{\bibinfo{volume}{69}},
  \bibinfo{pages}{045313} (\bibinfo{year}{2004}).

\bibitem[{\citenamefont{Bockhorn et~al.}(2011)\citenamefont{Bockhorn, Barthold,
  Schuh, Wegscheider, and Haug}}]{BockhornPRB2011}
\bibinfo{author}{\bibfnamefont{L.}~\bibnamefont{Bockhorn}},
  \bibinfo{author}{\bibfnamefont{P.}~\bibnamefont{Barthold}},
  \bibinfo{author}{\bibfnamefont{D.}~\bibnamefont{Schuh}},
  \bibinfo{author}{\bibfnamefont{W.}~\bibnamefont{Wegscheider}},
  \bibnamefont{and} \bibinfo{author}{\bibfnamefont{R.~J.} \bibnamefont{Haug}},
  \bibinfo{journal}{Phys. Rev. B} \textbf{\bibinfo{volume}{83}},
  \bibinfo{pages}{113301} (\bibinfo{year}{2011}).

\bibitem[{\citenamefont{Kozikov et~al.}(2010)\citenamefont{Kozikov, Savchenko,
  Narozhny, and Shytov}}]{Kozikov2010}
\bibinfo{author}{\bibfnamefont{A.~A.} \bibnamefont{Kozikov}},
  \bibinfo{author}{\bibfnamefont{A.~K.} \bibnamefont{Savchenko}},
  \bibinfo{author}{\bibfnamefont{B.~N.} \bibnamefont{Narozhny}},
  \bibnamefont{and} \bibinfo{author}{\bibfnamefont{A.~V.}
  \bibnamefont{Shytov}}, \bibinfo{journal}{Phys. Rev. B}
  \textbf{\bibinfo{volume}{82}}, \bibinfo{pages}{075424}
  (\bibinfo{year}{2010}).

\bibitem[{\citenamefont{Jouault et~al.}(2011)\citenamefont{Jouault, Jabakhanji,
  Camara, Desrat, Consejo, and Camassel}}]{JouaultPRB2011}
\bibinfo{author}{\bibfnamefont{B.}~\bibnamefont{Jouault}},
  \bibinfo{author}{\bibfnamefont{B.}~\bibnamefont{Jabakhanji}},
  \bibinfo{author}{\bibfnamefont{N.}~\bibnamefont{Camara}},
  \bibinfo{author}{\bibfnamefont{W.}~\bibnamefont{Desrat}},
  \bibinfo{author}{\bibfnamefont{C.}~\bibnamefont{Consejo}}, \bibnamefont{and}
  \bibinfo{author}{\bibfnamefont{J.}~\bibnamefont{Camassel}},
  \bibinfo{journal}{Phys. Rev. B} \textbf{\bibinfo{volume}{83}},
  \bibinfo{pages}{195417} (\bibinfo{year}{2011}).

\bibitem[{\citenamefont{Lara-Avila et~al.}(2011)\citenamefont{Lara-Avila,
  Tzalenchuk, Kubatkin, Yakimova, Janssen, Cedergren, Bergsten, and
  Fal'ko}}]{Lara-AvilaPRL2011}
\bibinfo{author}{\bibfnamefont{S.}~\bibnamefont{Lara-Avila}},
  \bibinfo{author}{\bibfnamefont{A.}~\bibnamefont{Tzalenchuk}},
  \bibinfo{author}{\bibfnamefont{S.}~\bibnamefont{Kubatkin}},
  \bibinfo{author}{\bibfnamefont{R.}~\bibnamefont{Yakimova}},
  \bibinfo{author}{\bibfnamefont{T.~J. B.~M.} \bibnamefont{Janssen}},
  \bibinfo{author}{\bibfnamefont{K.}~\bibnamefont{Cedergren}},
  \bibinfo{author}{\bibfnamefont{T.}~\bibnamefont{Bergsten}}, \bibnamefont{and}
  \bibinfo{author}{\bibfnamefont{V.}~\bibnamefont{Fal'ko}},
  \bibinfo{journal}{Phys. Rev. Lett.} \textbf{\bibinfo{volume}{107}},
  \bibinfo{pages}{166602} (\bibinfo{year}{2011}).

\bibitem[{\citenamefont{Jobst et~al.}(2012)\citenamefont{Jobst, Waldmann,
  Gornyi, Mirlin, and Weber}}]{JobstPRL2012}
\bibinfo{author}{\bibfnamefont{J.}~\bibnamefont{Jobst}},
  \bibinfo{author}{\bibfnamefont{D.}~\bibnamefont{Waldmann}},
  \bibinfo{author}{\bibfnamefont{I.~V.} \bibnamefont{Gornyi}},
  \bibinfo{author}{\bibfnamefont{A.~D.} \bibnamefont{Mirlin}},
  \bibnamefont{and} \bibinfo{author}{\bibfnamefont{H.~B.} \bibnamefont{Weber}},
  \bibinfo{journal}{Phys. Rev. Lett.} \textbf{\bibinfo{volume}{108}},
  \bibinfo{pages}{106601} (\bibinfo{year}{2012}).

\bibitem[{\citenamefont{Tikhonenko et~al.}(2009)\citenamefont{Tikhonenko,
  Kozikov, Savchenko, and Gorbachev}}]{TikhonenkoPRL2009}
\bibinfo{author}{\bibfnamefont{F.~V.} \bibnamefont{Tikhonenko}},
  \bibinfo{author}{\bibfnamefont{A.~A.} \bibnamefont{Kozikov}},
  \bibinfo{author}{\bibfnamefont{A.~K.} \bibnamefont{Savchenko}},
  \bibnamefont{and} \bibinfo{author}{\bibfnamefont{R.~V.}
  \bibnamefont{Gorbachev}}, \bibinfo{journal}{Phys. Rev. Lett.}
  \textbf{\bibinfo{volume}{103}}, \bibinfo{pages}{226801}
  (\bibinfo{year}{2009}).

\bibitem[{\citenamefont{Tikhonenko et~al.}(2008)\citenamefont{Tikhonenko,
  Horsell, Gorbachev, and Savchenko}}]{TikhonenkoPRL2008}
\bibinfo{author}{\bibfnamefont{F.~V.} \bibnamefont{Tikhonenko}},
  \bibinfo{author}{\bibfnamefont{D.~W.} \bibnamefont{Horsell}},
  \bibinfo{author}{\bibfnamefont{R.~V.} \bibnamefont{Gorbachev}},
  \bibnamefont{and} \bibinfo{author}{\bibfnamefont{A.~K.}
  \bibnamefont{Savchenko}}, \bibinfo{journal}{Phys. Rev. Lett.}
  \textbf{\bibinfo{volume}{100}}, \bibinfo{pages}{056802}
  (\bibinfo{year}{2008}), \bibinfo{note}{see also the supplementary material of
  this article.}

\bibitem[{\citenamefont{Baker et~al.}(2012)\citenamefont{Baker,
  Alexander-Webber, Altebaeumer, Janssen, Tzalenchuk, Lara-Avila, Kubatkin,
  Yakimova, Lin, Li et~al.}}]{BakerPRB2012}
\bibinfo{author}{\bibfnamefont{A.~M.~R.} \bibnamefont{Baker}},
  \bibinfo{author}{\bibfnamefont{J.~A.} \bibnamefont{Alexander-Webber}},
  \bibinfo{author}{\bibfnamefont{T.}~\bibnamefont{Altebaeumer}},
  \bibinfo{author}{\bibfnamefont{T.~J. B.~M.} \bibnamefont{Janssen}},
  \bibinfo{author}{\bibfnamefont{A.}~\bibnamefont{Tzalenchuk}},
  \bibinfo{author}{\bibfnamefont{S.}~\bibnamefont{Lara-Avila}},
  \bibinfo{author}{\bibfnamefont{S.}~\bibnamefont{Kubatkin}},
  \bibinfo{author}{\bibfnamefont{R.}~\bibnamefont{Yakimova}},
  \bibinfo{author}{\bibfnamefont{C.-T.} \bibnamefont{Lin}},
  \bibinfo{author}{\bibfnamefont{L.-J.} \bibnamefont{Li}},
  \bibnamefont{et~al.}, \bibinfo{journal}{Phys. Rev. B}
  \textbf{\bibinfo{volume}{86}}, \bibinfo{pages}{235441}
  (\bibinfo{year}{2012}).

\bibitem[{\citenamefont{Girvin et~al.}(1982)\citenamefont{Girvin, Jonson, and
  Lee}}]{GirvinPRB1982}
\bibinfo{author}{\bibfnamefont{S.~M.} \bibnamefont{Girvin}},
  \bibinfo{author}{\bibfnamefont{M.}~\bibnamefont{Jonson}}, \bibnamefont{and}
  \bibinfo{author}{\bibfnamefont{P.~A.} \bibnamefont{Lee}},
  \bibinfo{journal}{Phys. Rev. B} \textbf{\bibinfo{volume}{26}},
  \bibinfo{pages}{1651} (\bibinfo{year}{1982}).

\bibitem[{\citenamefont{Tanabe et~al.}(2010)\citenamefont{Tanabe, Sekine,
  Kageshima, Nagase, and Hibino}}]{TanabeAPEX2010}
\bibinfo{author}{\bibfnamefont{S.}~\bibnamefont{Tanabe}},
  \bibinfo{author}{\bibfnamefont{Y.}~\bibnamefont{Sekine}},
  \bibinfo{author}{\bibfnamefont{H.}~\bibnamefont{Kageshima}},
  \bibinfo{author}{\bibfnamefont{M.}~\bibnamefont{Nagase}}, \bibnamefont{and}
  \bibinfo{author}{\bibfnamefont{H.}~\bibnamefont{Hibino}},
  \bibinfo{journal}{Appl. Phys. Express} \textbf{\bibinfo{volume}{3}},
  \bibinfo{pages}{075102} (\bibinfo{year}{2010}).

\bibitem[{\citenamefont{Novoselov et~al.}(2006)\citenamefont{Novoselov, McCann,
  Morozov, Fal'ko, Katsnelson, Zeitler, Jiang, Schedin, and
  Geim}}]{Novoselov2006}
\bibinfo{author}{\bibfnamefont{K.~C.} \bibnamefont{Novoselov}},
  \bibinfo{author}{\bibfnamefont{E.}~\bibnamefont{McCann}},
  \bibinfo{author}{\bibfnamefont{S.~V.} \bibnamefont{Morozov}},
  \bibinfo{author}{\bibfnamefont{V.~I.} \bibnamefont{Fal'ko}},
  \bibinfo{author}{\bibfnamefont{M.~I.} \bibnamefont{Katsnelson}},
  \bibinfo{author}{\bibfnamefont{U.}~\bibnamefont{Zeitler}},
  \bibinfo{author}{\bibfnamefont{D.}~\bibnamefont{Jiang}},
  \bibinfo{author}{\bibfnamefont{F.}~\bibnamefont{Schedin}}, \bibnamefont{and}
  \bibinfo{author}{\bibfnamefont{A.~K.} \bibnamefont{Geim}},
  \bibinfo{journal}{Nat. Phys.} \textbf{\bibinfo{volume}{2}},
  \bibinfo{pages}{177} (\bibinfo{year}{2006}).

\bibitem[{\citenamefont{Tanabe et~al.}(2011)\citenamefont{Tanabe, Sekine,
  Kageshima, Nagase, and Hibino}}]{TanabePRB2011}
\bibinfo{author}{\bibfnamefont{S.}~\bibnamefont{Tanabe}},
  \bibinfo{author}{\bibfnamefont{Y.}~\bibnamefont{Sekine}},
  \bibinfo{author}{\bibfnamefont{H.}~\bibnamefont{Kageshima}},
  \bibinfo{author}{\bibfnamefont{M.}~\bibnamefont{Nagase}}, \bibnamefont{and}
  \bibinfo{author}{\bibfnamefont{H.}~\bibnamefont{Hibino}},
  \bibinfo{journal}{Phys. Rev. B} \textbf{\bibinfo{volume}{84}},
  \bibinfo{pages}{115458} (\bibinfo{year}{2011}).

\bibitem[{\citenamefont{Ponomarenko et~al.}(2004)\citenamefont{Ponomarenko,
  de~Lang, de~Visser, Kulbachinskii, Galiev, Künzel, and
  Pruisken}}]{PonomarenkoSSComm2004}
\bibinfo{author}{\bibfnamefont{L.}~\bibnamefont{Ponomarenko}},
  \bibinfo{author}{\bibfnamefont{D.}~\bibnamefont{de~Lang}},
  \bibinfo{author}{\bibfnamefont{A.}~\bibnamefont{de~Visser}},
  \bibinfo{author}{\bibfnamefont{V.}~\bibnamefont{Kulbachinskii}},
  \bibinfo{author}{\bibfnamefont{G.}~\bibnamefont{Galiev}},
  \bibinfo{author}{\bibfnamefont{H.}~\bibnamefont{Künzel}}, \bibnamefont{and}
  \bibinfo{author}{\bibfnamefont{A.}~\bibnamefont{Pruisken}},
  \bibinfo{journal}{Solid State Commun.} \textbf{\bibinfo{volume}{130}},
  \bibinfo{pages}{705 } (\bibinfo{year}{2004}).

\bibitem[{\citenamefont{Karmakar et~al.}(2004)\citenamefont{Karmakar, Gokhale,
  Shah, Arora, de~Lang, de~Visser, Ponomarenko, and
  Pruisken}}]{KarmakarPhysE2004}
\bibinfo{author}{\bibfnamefont{B.}~\bibnamefont{Karmakar}},
  \bibinfo{author}{\bibfnamefont{M.}~\bibnamefont{Gokhale}},
  \bibinfo{author}{\bibfnamefont{A.}~\bibnamefont{Shah}},
  \bibinfo{author}{\bibfnamefont{B.}~\bibnamefont{Arora}},
  \bibinfo{author}{\bibfnamefont{D.}~\bibnamefont{de~Lang}},
  \bibinfo{author}{\bibfnamefont{A.}~\bibnamefont{de~Visser}},
  \bibinfo{author}{\bibfnamefont{L.}~\bibnamefont{Ponomarenko}},
  \bibnamefont{and} \bibinfo{author}{\bibfnamefont{A.}~\bibnamefont{Pruisken}},
  \bibinfo{journal}{Physica E} \textbf{\bibinfo{volume}{24}},
  \bibinfo{pages}{187 } (\bibinfo{year}{2004}).

\bibitem[{Dru()}]{Drude}
\bibinfo{note}{$(W/L)\times R_{0}=h/(2e^{2}\tau_{0}v_{F}\sqrt{\pi n})$, see
  Refs.~\onlinecite{Lara-AvilaPRL2011,CN2009}}.

\bibitem[{\citenamefont{Das~Sarma et~al.}(2011)\citenamefont{Das~Sarma, Adam,
  Hwang, and Rossi}}]{SarmaRMP2011}
\bibinfo{author}{\bibfnamefont{S.}~\bibnamefont{Das~Sarma}},
  \bibinfo{author}{\bibfnamefont{S.}~\bibnamefont{Adam}},
  \bibinfo{author}{\bibfnamefont{E.~H.} \bibnamefont{Hwang}}, \bibnamefont{and}
  \bibinfo{author}{\bibfnamefont{E.}~\bibnamefont{Rossi}},
  \bibinfo{journal}{Rev. Mod. Phys.} \textbf{\bibinfo{volume}{83}},
  \bibinfo{pages}{407} (\bibinfo{year}{2011}).

\bibitem[{\citenamefont{Castro~Neto et~al.}(2009)\citenamefont{Castro~Neto,
  Guinea, Peres, Novoselov, and Geim}}]{CN2009}
\bibinfo{author}{\bibfnamefont{A.~H.} \bibnamefont{Castro~Neto}},
  \bibinfo{author}{\bibfnamefont{F.}~\bibnamefont{Guinea}},
  \bibinfo{author}{\bibfnamefont{N.~M.~R.} \bibnamefont{Peres}},
  \bibinfo{author}{\bibfnamefont{K.~S.} \bibnamefont{Novoselov}},
  \bibnamefont{and} \bibinfo{author}{\bibfnamefont{A.~K.} \bibnamefont{Geim}},
  \bibinfo{journal}{Rev. Mod. Phys.} \textbf{\bibinfo{volume}{81}},
  \bibinfo{pages}{109} (\bibinfo{year}{2009}).

\end{thebibliography}

\end{document}